\documentclass[modern]{aastex63}

\usepackage{xcolor}
\usepackage{booktabs}
\accepted{04/28/2021}
\submitjournal{ApJ}
\shorttitle{Flux ropes with field-aligned flows}
\shortauthors{Chen et al.} %%
\graphicspath{{FIGS/}}
\begin{document}

\title{Small-scale Magnetic Flux Ropes with Field-aligned Flows via the PSP In-situ Observations}

\correspondingauthor{Qiang Hu}
\email{qh0001@uah.edu}
% ----------------------------------------------------------

%\author[0000-0002-0786-7307]{Greg J. Schwarz}
%\affiliation{American Astronomical Society \\
%1667 K Street NW, Suite 800 \\

% ----------------------------------------------------------
\author[0000-0002-0065-7622]{Yu Chen}
\affiliation{Center for Space Plasma and Aeronomic Research (CSPAR), The University of Alabama in Huntsville, Huntsville, AL 35805, USA}

\author[0000-0002-7570-2301]{Qiang Hu}
\affiliation{Center for Space Plasma and Aeronomic Research (CSPAR), The University of Alabama in Huntsville, Huntsville, AL 35805, USA}
\affiliation{Department of Space Science, The University of Alabama in Huntsville, Huntsville, AL 35805, USA}

\author[0000-0002-4299-0490]{Lingling Zhao}
\affiliation{Center for Space Plasma and Aeronomic Research (CSPAR), The University of Alabama in Huntsville, Huntsville, AL 35805, USA}

\author[0000-0002-7077-930X]{Justin C. Kasper}
\affiliation{BWX Technologies, Inc., Washington, DC, 20002, USA}
\affiliation{Department of Climate and Space Sciences and Engineering, University of Michigan, Ann Arbor, MI 48109, USA}

\author[0000-0002-9954-4707]{Jia Huang}
\affiliation{Department of Climate and Space Sciences and Engineering, University of Michigan, Ann Arbor, MI 48109, USA}
% ----------------------------------------------------------

%\collaboration{1}{(AAS Journals Data Scientists collaboration)}
%
%\author{Butler Burton}
%\affiliation{Leiden University}
%\affiliation{AAS Journals Associate Editor-in-Chief}
%\nocollaboration{1}
%\altaffiliation{AASTeX v6+ programmer}
%\collaboration{1}{(LaTeX collaboration)}
%\affiliation{AAS Director of Publishing}
%\affiliation{American Astronomical Society \\
%1667 K Street NW, Suite 800 \\
%\author{Scott Chernoff}
%\affiliation{IOP Publishing, Washington, DC 20005}
%%% Note that the \and command from previous versions of AASTeX is now



% ----------------------------------------------------------
\begin{abstract}

Magnetic flux rope, formed by the helical magnetic field lines, can sometimes remain its shape while carrying significant plasma flow that is aligned with the local magnetic field. We report the existence of such structures and static flux ropes by applying the Grad-Shafranov-based algorithm to the Parker Solar Probe (PSP) in-situ measurements in the first five encounters. These structures are detected at heliocentric distances, ranging from 0.13 to 0.66 au, in a total of 4-month time period. We find that flux ropes with field-aligned flows, although occur more frequently, have certain properties similar to those of static flux ropes, such as the decaying relations of the magnetic fields within structures with respect to heliocentric distances. Moreover, these events are more likely with magnetic pressure dominating over the thermal pressure. About one-third of events are detected in the relatively fast-speed solar wind. Taking into account the high Alfv\'enicity, we also compare with switchback spikes identified during three encounters and interpret their inter-relations. We find that some switchbacks can be detected when the spacecraft traverses flux rope-like structures. The cross-section maps for selected events are presented via the new Grad-Shafranov type reconstruction. Finally, the possible evolution of the magnetic flux rope structures in the inner heliosphere is discussed. 
\end{abstract} 
% ----------------------------------------------------------


% ----------------------------------------------------------
% \keywords{solar wind --- turbulence --- magnetohydrodynamics (MHD)
% --- methods: data analysis}
%% and \citet commands to identify citations. The citations are

% -----------------------------
\section{Introduction}\label{sec:intro}

Magnetic flux rope is a structure in which the magnetic field lines are tangled up around a central axis with usually stronger axial magnetic field. Generally, it possesses very different temporal and spatial scales, including large-scale (also known as the magnetic cloud, \cite{Zurbuchen2006}) and small-scale flux ropes \citep{Moldwin2000}. However, observational analyses from 0.3 to 9 au reveal that their duration and scale sizes, although vary with heliocentric distances and heliographic latitudes, have rather continuous distributions \citep{Chen2019,Chen2020}. In a recent study \citep{Chen2020b}, we applied the Grad-Shafranov (GS)-based algorithm to the Parker Solar Probe (PSP) dataset in the first two encounters and reported over 40 small-scale flux ropes, while the monthly occurrence rate of the same type of structures at larger radial distances was found to be in the order of a few hundreds. Such a remarkable difference in event counts at different radial distances hints that these small-scale flux ropes may have multiple origins.

The GS-based algorithm is one of the applications of the GS reconstruction \citep{Sonnerup1996,Hau1999} to magnetic flux ropes. Such a reconstruction technique has been developed for several decades \citep{Sonnerup2006,Hu2017GSreview}. It treats the one-dimensional (1D) spacecraft measurements as direct input to an initial value problem and recovers the two-dimensional (2D) structure in quasi-static equilibrium by utilizing several field line invariants. The GS technique was initially used to determine the flux rope invariant axis and derive its 2D cross-section \citep{Hu2001,Hu2002,Hu2004}. Later, it was compiled into an automated program to identify these structures \citep{Zheng2017,Hu2018}. An extension of the GS reconstruction was proved theoretically to be able to govern a structure in which the plasma flow is significant and aligned with the magnetic field (\cite{Sonnerup2006} and references therein). The in-situ observation and this application were first reported in \cite{Teh2007}. Those authors reconstructed the magnetic field configuration surrounding the spacecraft paths when the two Cluster spacecraft crossed the magnetopause. Additional results indicated that the cross-section maps of magnetic field lines were nearly consistent with the ones obtained with the original magnetohydrostatic approximation \citep{Hasegawa2004}. However, the application of this GS-type formulation to in-situ spacecraft measurements in the solar wind is still rare.

The PSP spacecraft provides in-situ measurements at unprecedentedly close distances to the Sun. The initial findings in the first two encounters include a lot of structures that have simultaneous occurrences of the radial magnetic field reversals and velocity spikes \citep{Bale2019,Kasper2019}. Such structures are known as switchbacks. They were found to be highly Alfv\'enic and also appeared in the slow solar wind. The magnitude of the magnetic field and the proton number density within the event interval remain largely constant, while the Poynting flux and the thermal energy increase inside \citep{Mozer2020}. Moreover, the statistics reveal that switchbacks have a wide range of duration, i.e., from less than one second to a few hours, following a power-law distribution \citep{Dudok2020}. They also showed the possibility that these switchbacks may be detected when a spacecraft crosses the kinked portions of magnetic flux tubes (or ropes). 

Due to the high Alfv\'enicity, it may be improper to compare the switchback with the static flux rope directly. In the situation of a static flux rope, the remaining flow or velocity in the frame moving with the flux rope must vanish. Thus, candidates with significant flows were excluded in our previous studies. The traditional observational analysis of magnetic flux rope has to be extended to accommodate the Alfv\'enic structures. In the situation with non-vanishing flow, both the magnitude and direction of the flow are checked against the local magnetic field to detect the cases with modest to high Alfv\'enicity, as quantified by the Alfv\'en Mach number, $M_A$. 

If we consider that the definition of a magnetic flux rope mainly relies on twisted field lines, a flux rope with significant field-aligned flow (hereafter, FRFF) also meets the definition but maintains dynamic equilibrium with inertia force included. Therefore, what differs from the static flux rope (hereafter, FR) is the existence of this significant flow, which is nearly Alfv\'enic and can be either parallel or anti-parallel to the local magnetic field lines. Such a situation makes the Alfv\'enicity within a flux rope or flux rope-like structure non-negligible. Notice that these field-aligned flow structures were found or speculated to exist in many locations and take different forms, such as the so-called Alfv\'en vortices in the Earth's magnetosheath \citep{Alexandrova2006} and solar wind \citep{Roberts2016}, the torsional Alfv\'en waves \citep{Higginson2018}, etc. Recently, \cite{Shi2021} also reported the existence of Alfv\'en waves from PSP measurements that are inside a small-scale flux rope with ion-cyclotron waves located near the boundaries. Now, we can proceed to extend the detection of small-scale structures to close heliocentric distances and compare them with switchbacks. 

In this paper, we apply the GS-based automated algorithms to identify both FR and FRFF events by using the PSP in-situ measurements during the time periods around the first five perihelia, i.e., encounters 1 to 5 (E1-E5). Notice that the scales of these two types of structures in our current analysis range from several minutes to a few hours, which are much smaller than those of the magnetic clouds. This paper is organized in the following order. In Section \ref{sec:method}, we briefly introduce the GS-based detection and its application to FR and FRFF events. Also, we recap the new GS reconstruction method \citep{Teh2018}, which is implemented to recover the 2D cross-section maps of FR/FRFF events. The results of the fifth encounter are presented in Section \ref{sec:overview} as an example together with all detection results. In particular, we analyze the macroscopic properties of FRs and FRFFs. The basic bulk parameters versus the heliocentric distances, including the magnetic field, duration, scale size, etc., are presented via 2D histograms. In Section \ref{sec:switchbacks}, we compare the FR/FRFF event lists with the switchback spikes in a statistical manner, and present the cross-section maps for selected cases. The major findings and discussions are summarized in Section \ref{sec:sum}.

\section{Method and Data} \label{sec:method}

%%%% Method 

The GS-based automated algorithm was proposed in \cite{zhengandhu2018} and \cite{Hu2018} based on the approach of the GS reconstruction (see \cite{Hu2017GSreview} for a comprehensive review). Guaranteed by the standard GS equation, i.e., $\nabla^2 A = -\mu_0dP_t/dA = -\mu_0d(p+B_z^2/2\mu_0)/dA$, the transverse pressure $P_t$ is a single variable function of the magnetic flux function $A$ \citep{Sonnerup1996,Hau1999,Hu2001,Hu2002}. For a flux rope structure, a spacecraft would cross the same set of iso-surfaces of $A$ twice along the inbound and outbound paths, but in the reverse order for the latter. Therefore, the one-to-one correspondence between $P_t$ and $A$ as well as the folding of $P_t$ versus $A$ from the inbound portion over the outbound portion is expected. The identification of magnetic flux ropes is thus facilitated by examining whether there is any in-situ data interval corresponding to a time-series data array possessing a good double-folding pattern in $P_t(A)$. 

In the GS-based algorithm, we employ a set of sliding search windows from 5.6 to 360 minutes. Within each window, all parameters are transformed into a co-moving frame of reference, i.e., usually the de Hoffmann–Teller (HT) frame \citep{deHoffmann1950}. After calculating $P_t$ and $A$, we use two residues to examine whether a candidate has a double-folding pattern in $P_t(A)$ with good quality. The boundaries of an event are thus determined by the locations where such a pattern starts and ends. Also,  we apply limits to the field strength to remove small fluctuations (see the flowchart at \url{http://fluxrope.info/flowchart.html}, and \cite{Hu2018}, for more details). For the FRFF events, the same algorithm still applies with a slight modification to the form of $P_t(A)$, and an additional criterion to ensure that the flow is approximately field-aligned. 

The traditional GS-based algorithm calculates the Wal\'en test slope of each candidate interval and excludes events with relatively high Alfv\'enicity. Such a slope (largely equivalent to the average Alfv\'en Mach number of the remaining flow, $M_A$) is obtained by the linear regression between the remaining flow velocity in the HT frame, i.e., $\textbf{V}_{remaining}=\textbf{V}-\textbf{V}_{HT}$, and the local Alfv\'en velocity $\textbf{V}_{A}$, where the plasma flow velocity measured in the spacecraft frame is denoted by $\mathbf{V}$. In this study, events that have slopes less than 0.3 are defined as traditional static FRs based on our experiences \citep{Hu2018}, while those with slopes larger than 0.3 have significant remaining flows and are thus regarded as candidates for FRFFs. In order to meet the definition of field-aligned flow, we also require the correlation coefficient $R$ of these two velocities to be greater than 0.8.

The cross-section map of FRFF in this study is obtained through the new GS reconstruction technique by employing an extended GS-type equation. It is for a special case with an anisotropic plasma pressure or alternatively the case of isotropic pressure with field-aligned flow \citep{Teh2018}:
\begin{equation}
\nabla^2 A'=-\frac{1}{2}\frac{dF^2_z}{dA'}-\mu_0\frac{dp_T}{dA'}+\mu_0(p_T+\frac{F^2}{2\mu_0})\frac{d\ln(1-\alpha)}{dA'},
\label{eq:eq1}
\end{equation}

where for the magnetic induction $\mathbf{B}$, $\textbf{F}=(1-\alpha)\textbf{B}$,  
$A'(x,0)=-\int (1-\alpha)B_y(x',0) dx'$, and $p_T=(1-\alpha)p_\perp+\alpha(1-\alpha)^{-1}(F^2/2\mu_0)$, with $\alpha =\mu_0 (p_\parallel-p_\perp)/B^2$. Here, $A'$ is the modified magnetic flux function, and the perpendicular plasma pressure is denoted by $p_\perp$. The prime symbol is used to distinguish from the magnetic flux function $A$ in the original GS equation. It should be noted that for $\alpha\equiv 0$, the above GS-type equation reduces to the original GS equation.

As suggested in \cite{Teh2018}, the extended GS equation can be applied to the field-aligned flow by taking $\alpha = M_A^2$ and $p_\perp=p$, the isotropic pressure. In particular, the third term in Equation (\ref{eq:eq1}) vanishes for $\alpha\equiv Const$. By plugging in the expression of $p_T$, we obtain the new GS type equation applicable to the equilibrium with field-aligned flow of a constant proportionality factor $\alpha$, i.e., a constant Alfv\'en Mach number:
\begin{equation}
\label{eq:eq2}
\nabla^2 A'=-\mu_0\frac{d}{dA'}\left[(1-\alpha)^2\frac{B^2_z}{2\mu_0}+(1-\alpha)p+\alpha(1-\alpha)\frac{B^2}{2\mu_0}\right],
\end{equation}
where $A'$ remains unchanged and is related to the flux function $A=A'/(1-\alpha)$, $B_z$ is the axial magnetic field, and $p$ is the thermal pressure. Meanwhile, the double-folding requirement for the original GS reconstruction, in its most general form, becomes the new single-valued function relation, $P'_t~versus~A'$, where $P'_t=(1-\alpha)^2B_z^2/2\mu_0+(1-\alpha)p+\alpha(1-\alpha) B^2/2\mu_0$.

Since the data integrity of the magnetic field is usually better than that of the plasma bulk parameters, we simplify the detection algorithm by considering the axial magnetic pressure (the first term on the right hand of equation \ref{eq:eq2}) as the only contribution to $P'_t$, as a first step and for $\alpha=M_A^2\equiv Const$. This also results in minimal changes to our current search algorithm employing the original GS equation (in which $\alpha\equiv0$). While a similar approach with slight modification from the detection of static FRs is implemented for the detection of FRFFs, the new GS reconstruction (the full equation \ref{eq:eq2}) is applied to derive cross-section maps of FRFFs for $\alpha \ne 1.0$, as demonstrated in \cite{Teh2018}. The general rule of thumb is that the reconstruction can work independently from the detection procedures of magnetic flux ropes and similar structures. The latter can have more flexibility in terms of employing different sets of data products when data integrity issues (i.e., data gaps) arise.

The data in this paper are downloaded from the NASA CDAWeb with the tag "Only Good Quality". The magnetic field data, plasma bulk properties, and electron pitch angle distributions (ePADs) are provided by the FIELDS Experiment (\cite{Bale2016}), the Solar Wind Electrons Alphas and Protons (SWEAP; \cite{Kasper2016,Case2020}), and SWEAP SPAN-Electron Instrument \citep{Whittlesey2020}, respectively. Since the magnetic field and plasma data have different cadences, we downsample all datasets by averaging to a sample rate of 28 sec.

\section{Overview of Detection Results in the First Five PSP Encounters}\label{sec:overview}

\begin{table}
\begin{center}
\caption{FRs and FRFFs in the first five PSP encounters.}
\footnotesize
\begin{tabular}{lcccccc}
\toprule
PSP Encounter & E1 & E2 & E3 & E4 & E5 & Total\\
\midrule
Perihelion & Nov 5, 2018 & Apr 4, 2019 & Sep 1, 2019 & Jan 29, 2020 & June 7, 2020 & N/A\\
Duration (days) & 18 & 42 & 19 & 16 & 21 & 116\\
Distances (au) & 0.17-0.37 & 0.17-0.66 & 0.18-0.64& 0.13-0.37& 0.13-0.36& 0.13-0.66\\
FRs Count & 24 & 20 & 33 & 47 & 119 & 243\\
FRFFs Count & 262 & 339 & 189 & 215 & 148 & 1,153 \\
\bottomrule
\end{tabular}
\label{table:result}
\end{center}
\end{table}

We apply the aforementioned two GS-based detection algorithms to identify static FRs (see, e.g., \cite{Chen2020b}) and dynamic FRFFs during a number of days around the perihelion in each encounter. Table \ref{table:result} lists the dates of perihelia, duration of detection periods, ranges of the heliocentric distances, and total counts of static FRs and FRFF events respectively. The detection periods in E1, E2, and E3 are centered on the perihelia with extending periods to inbound or outbound spacecraft paths, while those in E4 \& E5 are completely centered on the perihelia. In these 5 encounters, we discovered a total of 243 FRs and 1,153 FRFFs in about a 4-month detection period. These records locate at heliocentric distances ranging from 0.13 to 0.66 au. Although the situation varies, FRFFs are generally detected more frequently than FRs. In particular, the event counts of FRFFs in E1 and E2 are about a dozen times more than those of FRs. This prominent difference can be ascribed to the prevalent existence of the Alfv\'enic structures in the slow solar wind near the perihelion. E3 also has a similar result, albeit the difference decreases due to large data gaps in the outbound path. Because of the heliospheric current sheet (HCS) crossings on 2020 Jan 20 and Feb 1 \citep[e.g.,][]{Zhao2020aa} in the E4, 36 FRs occur within 3 days after such structures, while the total count of FRFFs reduces a little due to data gaps (about 3 days) around perihelion. These gaps also occur in the E5 and result in a reduction of the event count of FRFF. Meanwhile, there are more FRs detected in this encounter, which is thus quite comparable to that of FRFFs.

\begin{figure}
\centering
\includegraphics[width=1.0\textwidth]{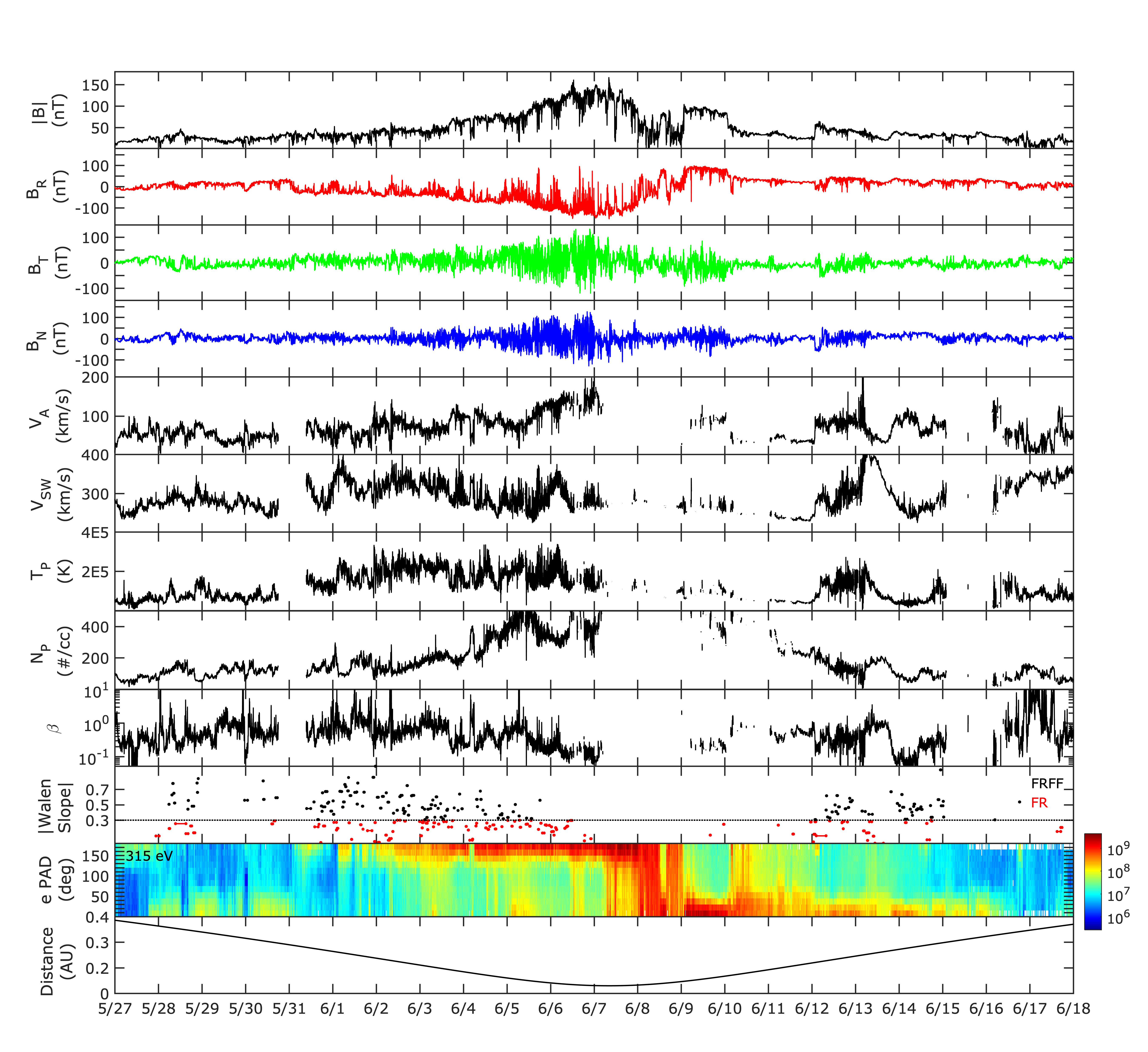}
\caption{Time-series plot in the E5 from 2020 May 27 to June 18. Panels from the top to bottom are the magnetic field and its three components in the RTN coordinates, the Alfv\'en speed, the solar wind bulk speed, the proton temperature, the proton number density, the plasma $\beta$, the absolute values of the Wal\'en test slopes with event duration indicated by lines between two dots, the electron pitch angle distribution (ePAD) at 315 eV energy channel, and the spacecraft radial distance from the Sun. The horizontal dotted line in the 10th panel marks the threshold value 0.3. Two groups of FR and FRFF events are separated by the red and black colors, respectively.}\label{fig:overview}
\end{figure}

Figure \ref{fig:overview} presents an overview of the elementary parameters in the PSP E5 from 2020 May 27 to June 18. During this time period, we detected 119 FRs and 148 FRFFs from 0.13 to 0.36 au. The starting and ending times of each event are marked by dots in the 10th panel, and lines are drawn between each pair of dots representing the duration. The red and black colors indicate events from groups of FRFF and FR respectively. They are separated by the horizontal dotted line, which marks the Wal\'en test threshold value 0.3. Moreover, most FRFFs possess slopes below the value 0.7. This tendency demonstrates that most FRFFs possess significant field-aligned flows, but may still deviate from pure Alfv\'en waves. The first four panels show the magnetic field measurements. As the PSP approached the perihelion, the magnitudes of the magnetic field and $B_R$ increase, while $B_T$ and $B_N$ components fluctuate more intensively. 

Notice that the $B_R$ has a shift from the negative peak (around 2020 June 8) to the positive values (2020 June 9). This sudden change in the radial magnetic field suggests the possible crossing of the HCS. Meanwhile, the 11th panel, i.e., the ePAD at 315 eV, also complies with this suggestion. The ePAD shows a prominent change of flux enhancement from 180$^\circ$ to 0$^\circ$, coinciding with the switch of the magnetic field polarities. These two angles indicate that the suprathermal electrons travel in a direction either parallel or anti-parallel to the magnetic field. As proposed in the previous study, e.g., \cite{Hu2018}, it is more likely for FRs to be detected in ambient solar wind near the HCS crossings. The cluster of FR records is quite evident before this structure, whereas it is unknown behind because of the data gaps.

Moreover, both $B_T$ and $B_N$ have sudden jumps in magnitudes between 2020 June 12 01:00 and 02:00, while $B_R$ has a change from 20 nT to -20 nT. These sharp changes are accompanied by increasing $V_{SW}$ and $T_p$ as well as a minor variation in the $N_p$. Such signatures may illustrate that a shock or a compressional wave occurs during that time period. As indicated in the panel of the Wal\'en test slopes, clustered FRs and FRFFs arise downstream of this transient structure. 

\begin{figure}
\centering
\includegraphics[width=.45\textwidth]{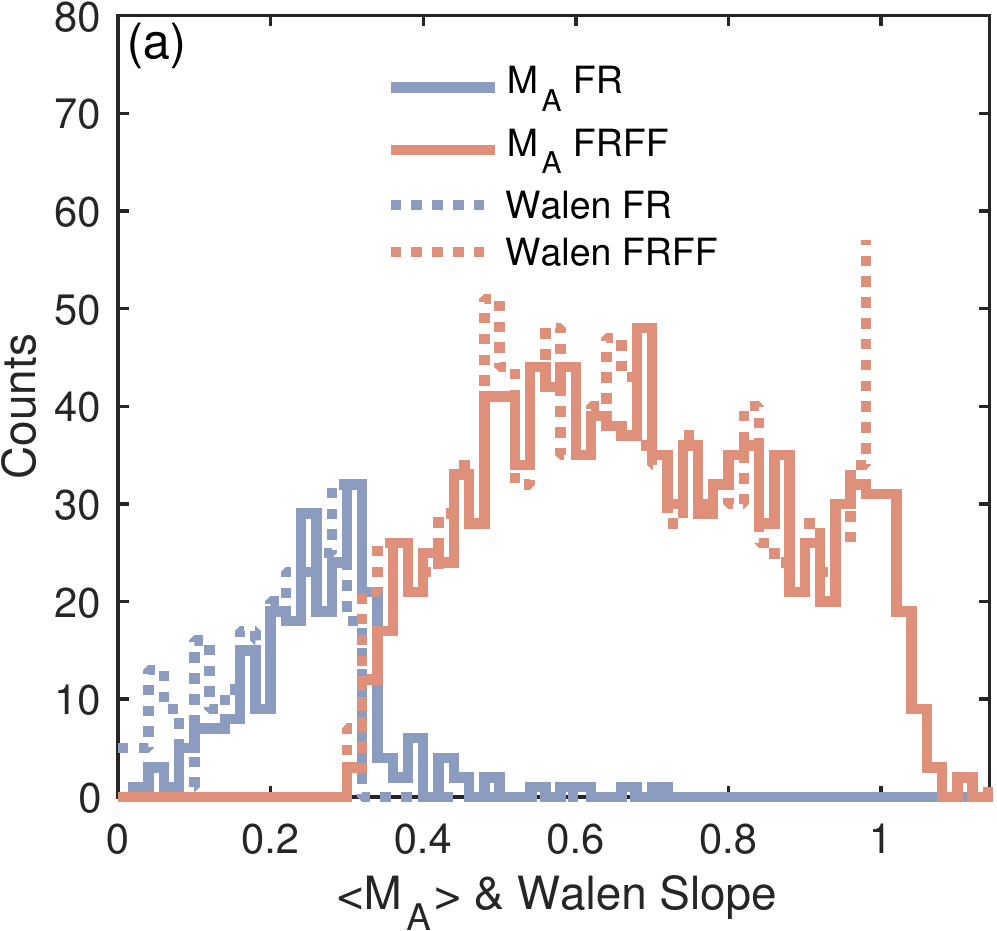}
\includegraphics[width=.45\textwidth]{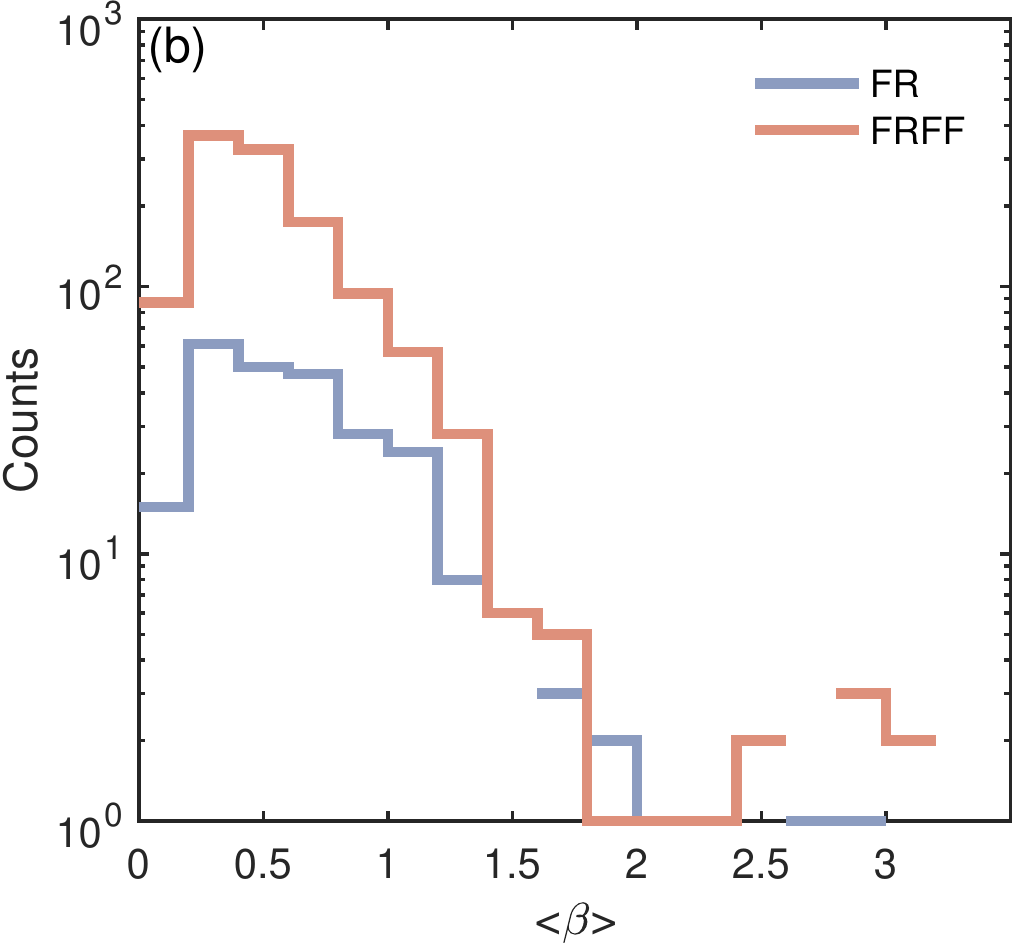}
\caption{Distributions of parameters for the FR and FRFF events in the first 5 PSP encounters: (a) the average Alfv\'en Mach number $\langle M_A\rangle$ and the absolute value of the Wal\'en test slope, and (b) plasma $\beta$. See legends for the parameters associated with each group.}
\label{fig:FRFF}
\end{figure}

We treat FRs and FRFFs in the first 5 PSP encounters as two individual groups. Figure \ref{fig:FRFF}(a) presents the distribution of the average Alfv\'en Mach number $\langle M_A\rangle$ and the absolute value of the Wal\'en test slope. Both parameters distinguish clearly in the two groups with 0.3 as a threshold value. Such a separation is predictable since these events are categorized based on the Wal\'en test slope, which is derived in a way analogous to $M_A$. Although there are some FRs possessing $\langle M_A\rangle$ greater than 0.3, which is probably due to extreme values within shorter event intervals, most events have $\langle M_A\rangle$ smaller than 0.3, i.e., corresponding to the static structures. Again, our event selection is based on the threshold value of the Wal\'en test slope, instead of $\langle M_A\rangle$. On the other hand, almost all FRFFs have $\langle M_A\rangle$ greater than 0.3. This inclination confirms that FRFFs are more Alfv\'enic when compared with the static FRs. Notice that some FRFFs even have $\langle M_A\rangle$ larger than 1 that hints the super-Alfv\'encity, while most structures have sub-Alfv\'enic flow. Figure \ref{fig:FRFF}(b) presents the plasma $\beta$ distributions, in which both groups tend to peak at $\beta\approx0.5$.

\begin{figure}
\centering
\includegraphics[width=.45\textwidth]{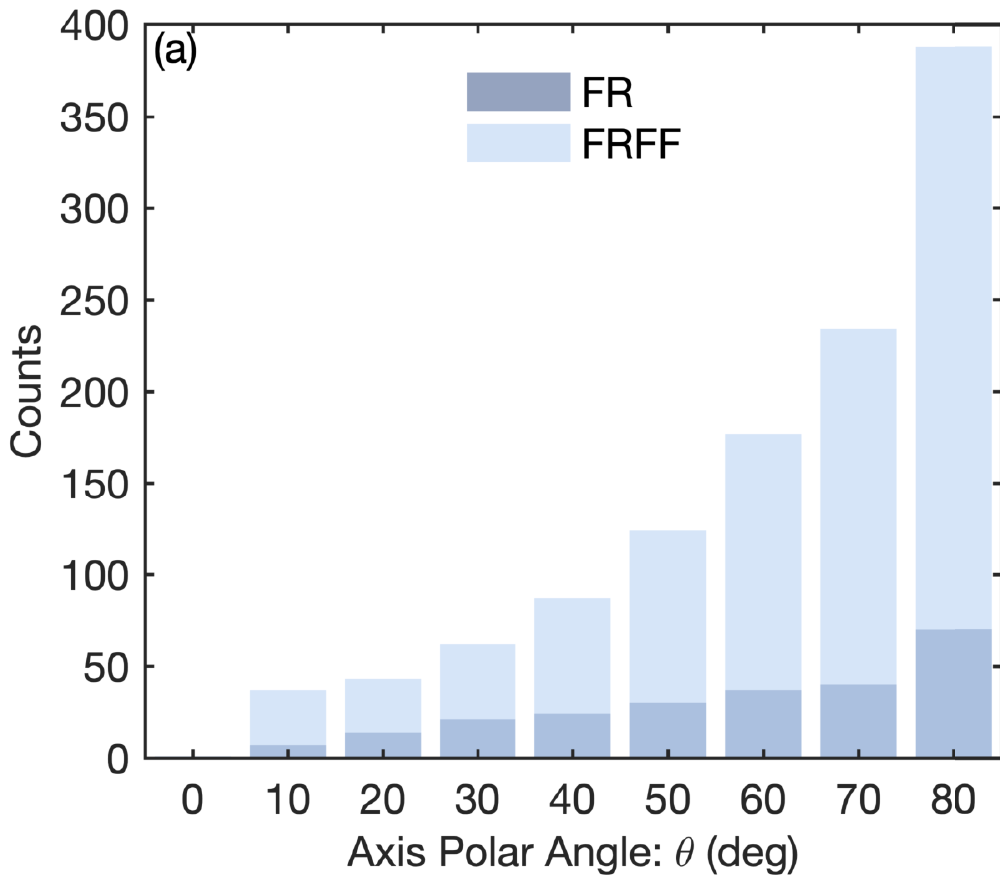}
\includegraphics[width=.45\textwidth]{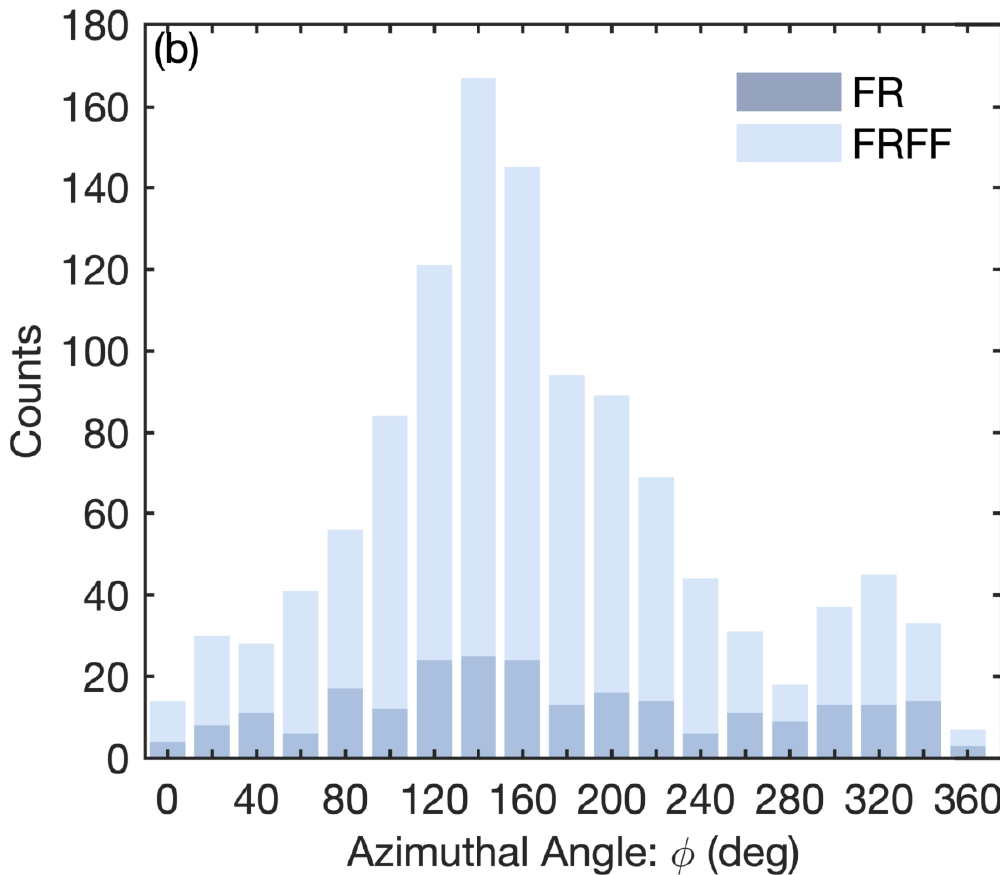}
\caption{Distributions of the axis orientations for the FR and FRFF events in the first five PSP encounters: (a) the polar angle $\theta$, and (b) the azimuthal angle $\phi$. }\label{fig:pr2}
\end{figure}

Figure \ref{fig:pr2} presents the distributions of the orientation angles of the invariant $z$-axis for both FR and FRFF events. The azimuthal angle $\phi$ is defined as the angle between R direction and projection of $z$-axis onto the RT-plane, while the polar angle $\theta$ is calculated between the N direction and the $z$-axis. Since the event counts of FR and FRFF are different, the group of FRFF tends to have a clear tendency that peaks at about $\phi\approx$ 140$^\circ$. Both groups of events tend to lie on the RT-plane.

\begin{figure}
\centering
\includegraphics[width=.3\textwidth]{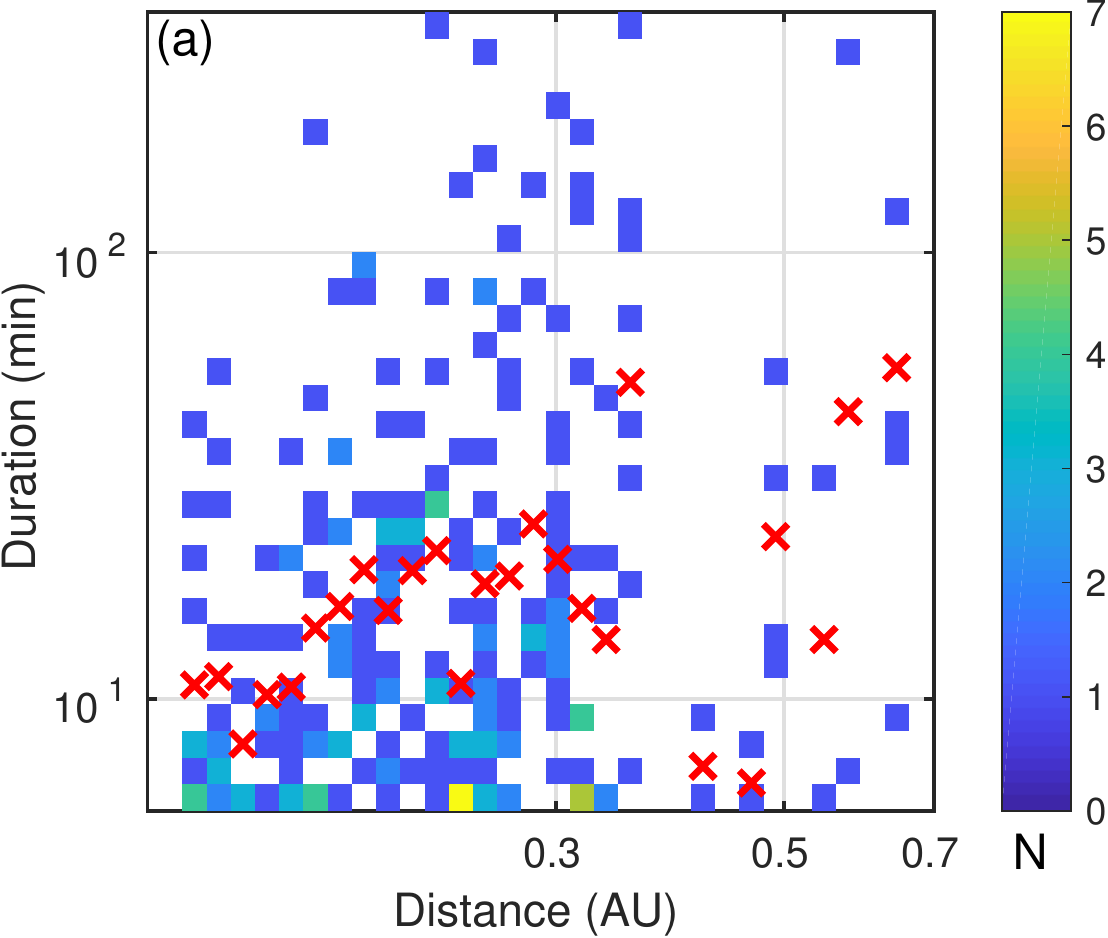}
\includegraphics[width=.308\textwidth]{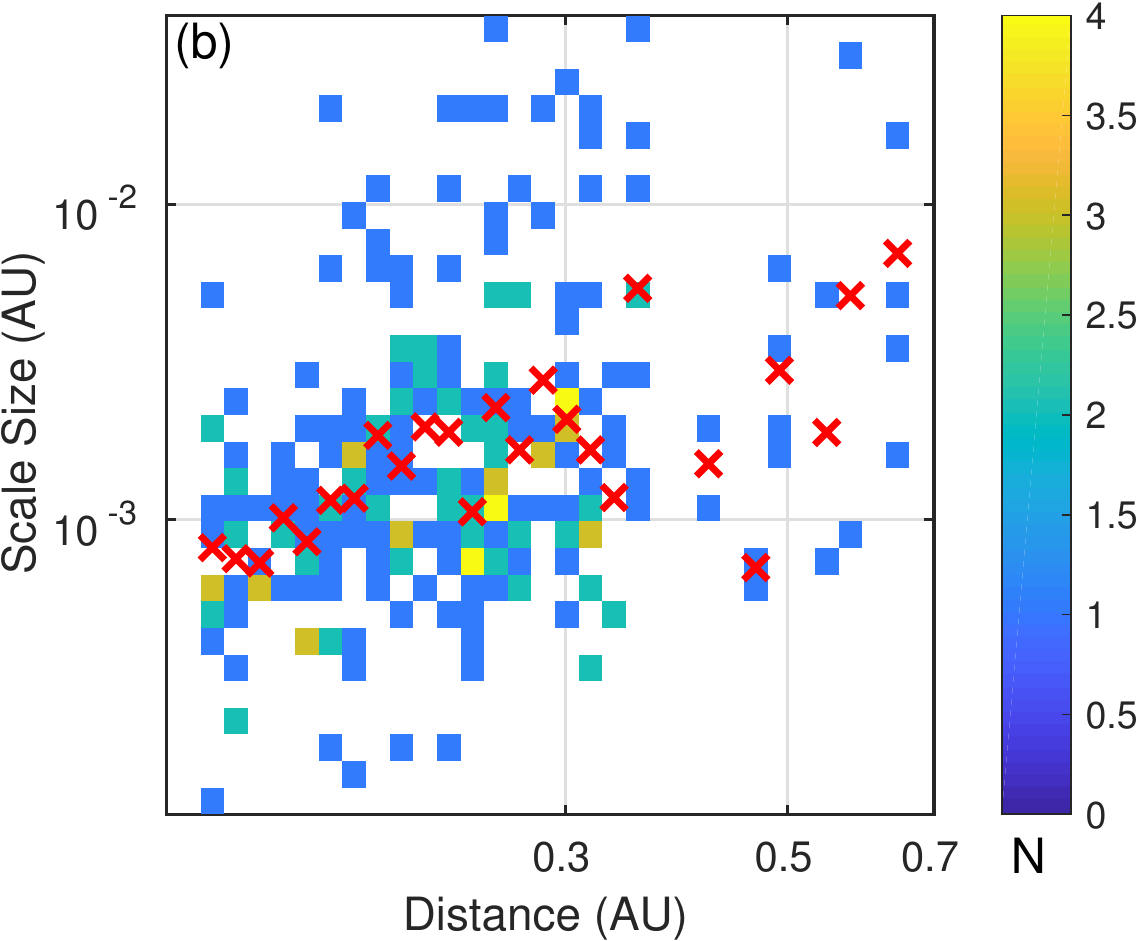}
\includegraphics[width=.3\textwidth]{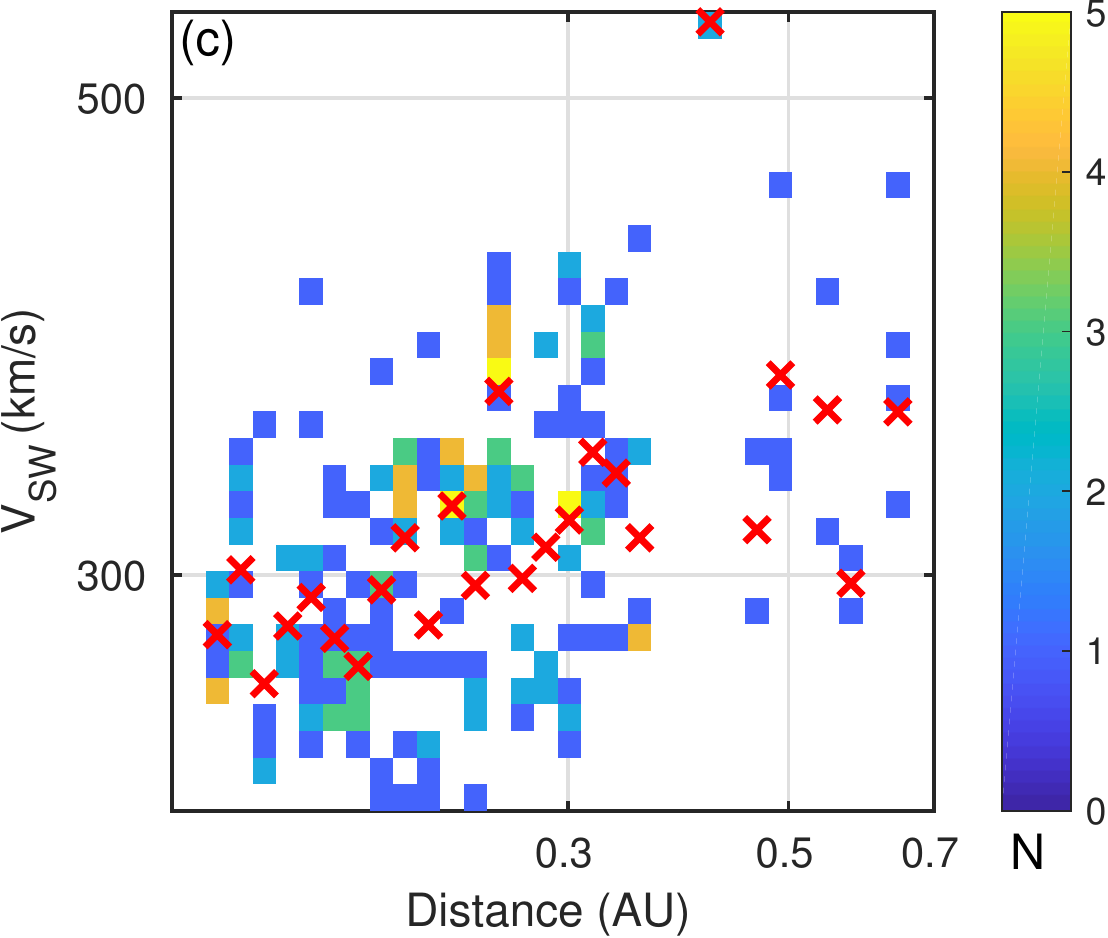}
\includegraphics[width=.3\textwidth]{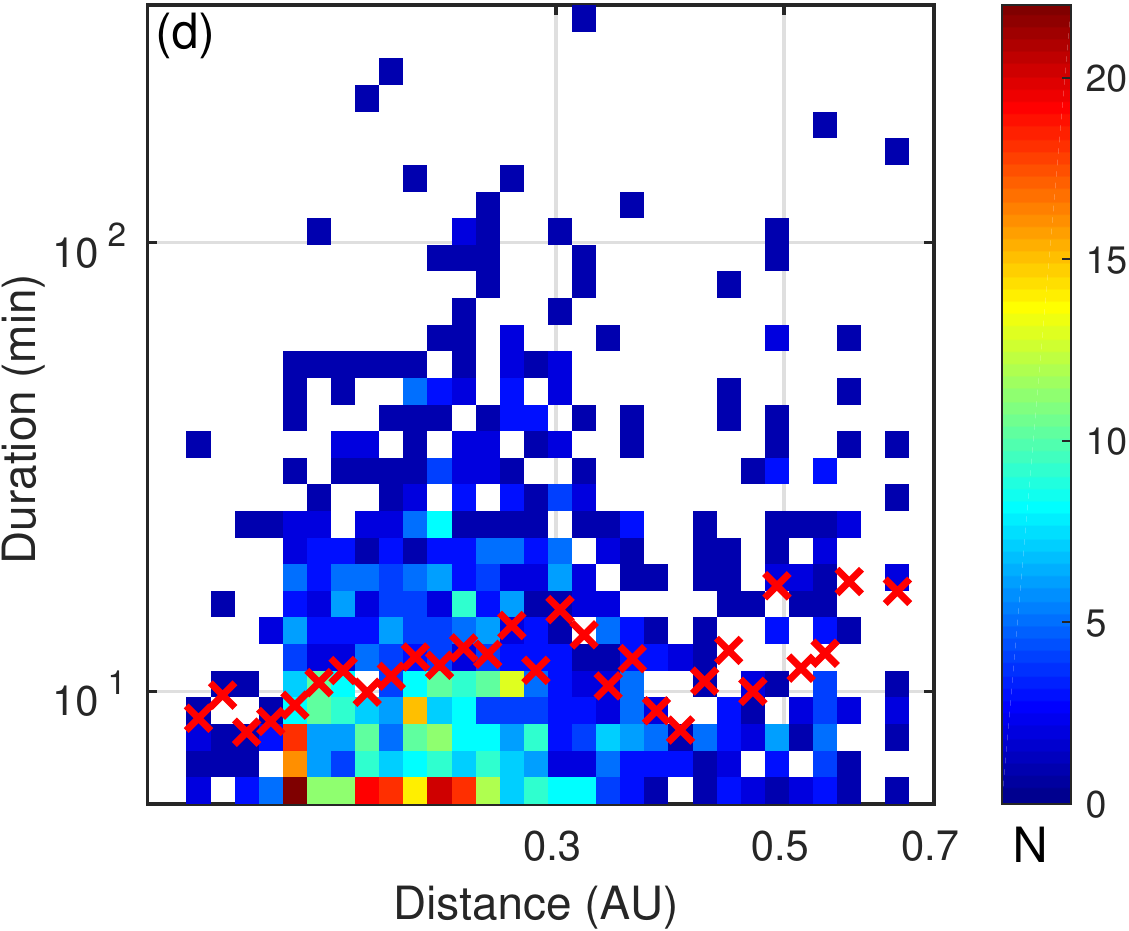}
\includegraphics[width=.3\textwidth]{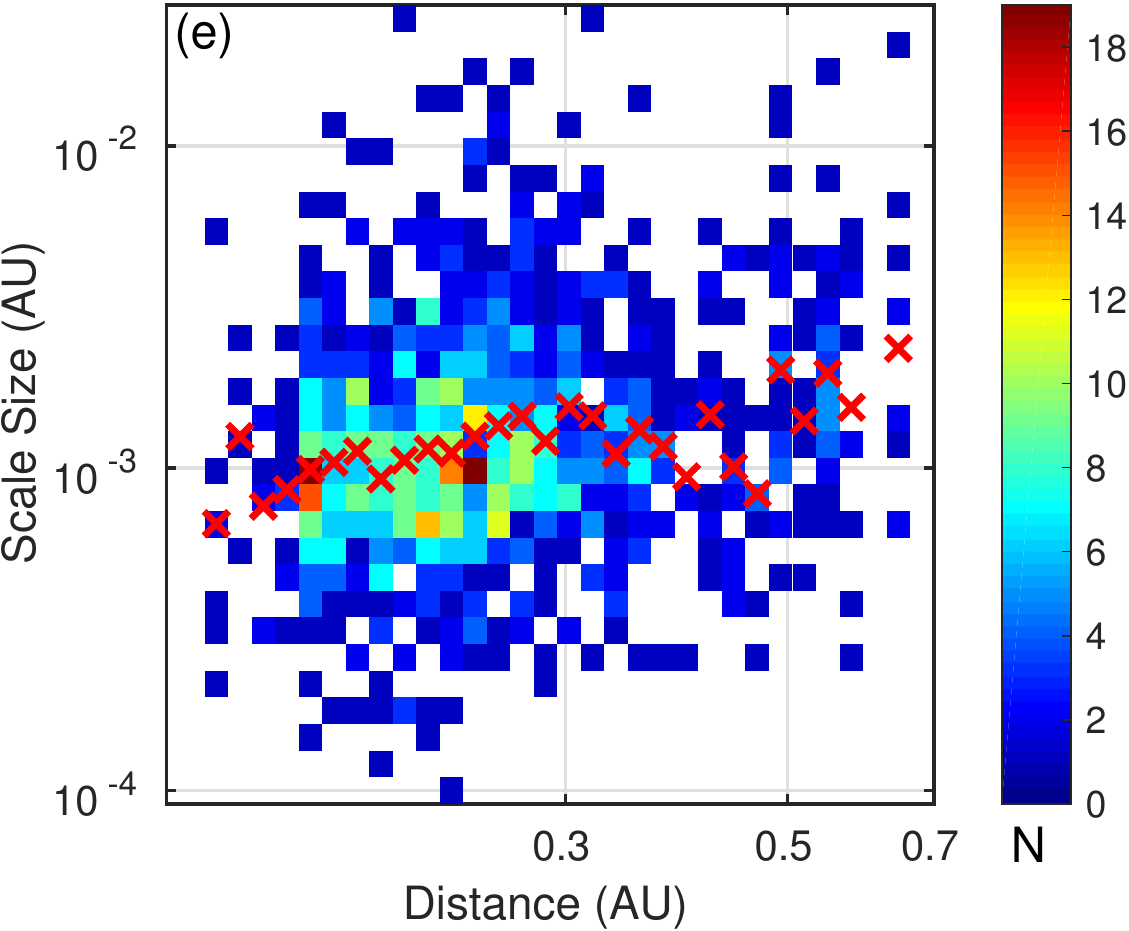}
\includegraphics[width=.3\textwidth]{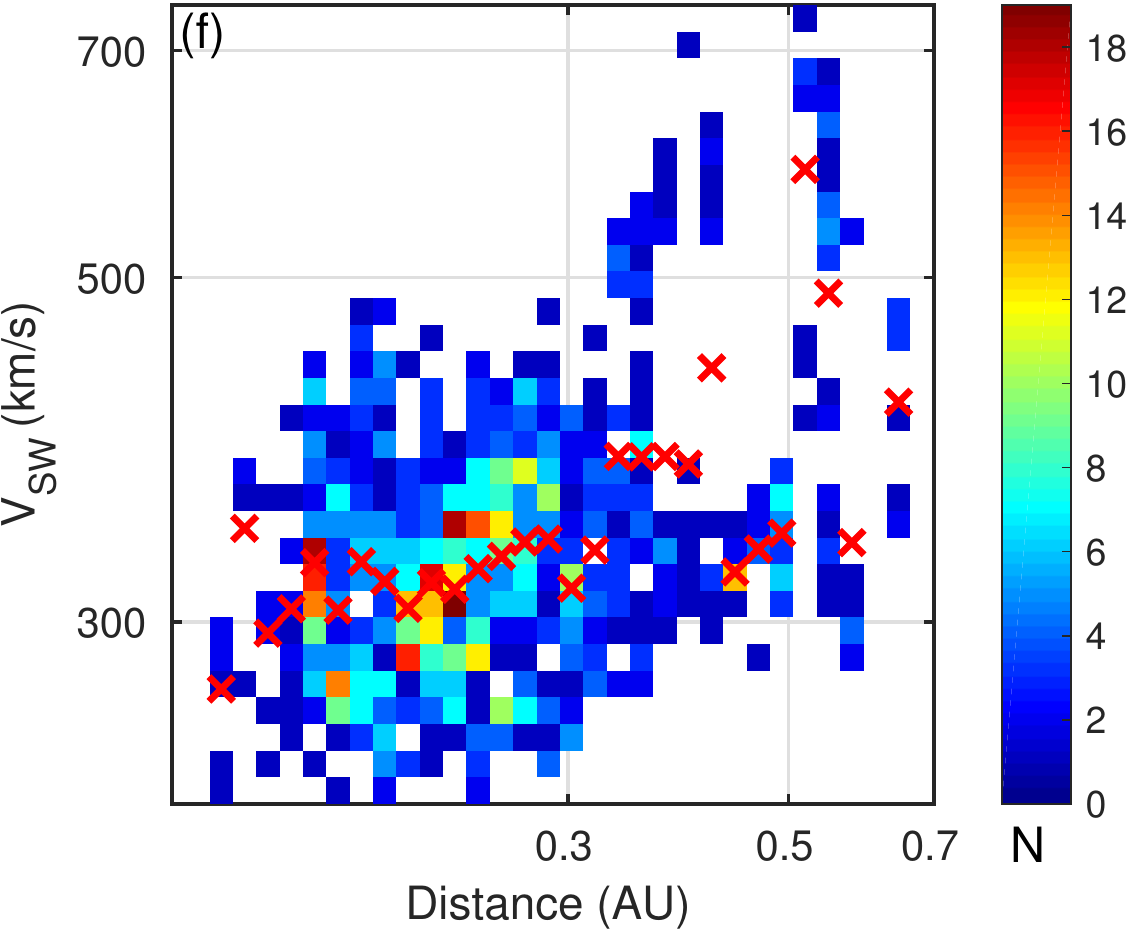}
\caption{The 2D distributions of properties of static FRs (the first row) and FRFFs  (the second row) with respect to the radial distances $r$: (a) \& (d) duration, (b) \& (e) scale size, and (c) \& (f) the averaged solar wind speed. The bin grids are 30 $\times$ 30 in size. The average values in each bin of $r$ are marked by red crosses. }\label{fig:FRFF1}
\end{figure}

Figure \ref{fig:FRFF1} shows the 2D histograms of event duration, scale size, and averaged solar wind speed. Both groups have events clustering around heliocentric distances less than 0.36 $\sim$ 0.37 au due to the available time periods of detection. Events with smaller duration dominate in both groups, especially for FRFF records. Similarly, most cases have scale sizes of less than 0.001 au. For FR events, these two parameters seem to distribute more randomly from 0.13 to 0.37 au due to small sample size. In comparison, FRFF events show tighter distributions with little change in duration and scale size over the range of radial distances. Such slow variation may demonstrate that the expansion in the radial direction is not significant at these heliocentric distances. Figure \ref{fig:FRFF1}(c) presents the distribution of the solar wind speed averaged in each event interval. Both fast- and slow-speed wind were observed by the PSP \citep{Kasper2019}, and both sets of events are detected in solar wind streams with variable speeds. Static FRs are detected frequently in the slow solar wind near the perihelion. The average solar wind speed of this group is 315 km s$^{-1}$, and only 17\% of events have average speeds larger than 350 km s$^{-1}$. This ratio is about 33\% in the group of FRFFs, which indicates that they have relatively more records in the medium and fast-speed winds when comparing with FRs.

\begin{figure}
\centering
\includegraphics[width=.3\textwidth]{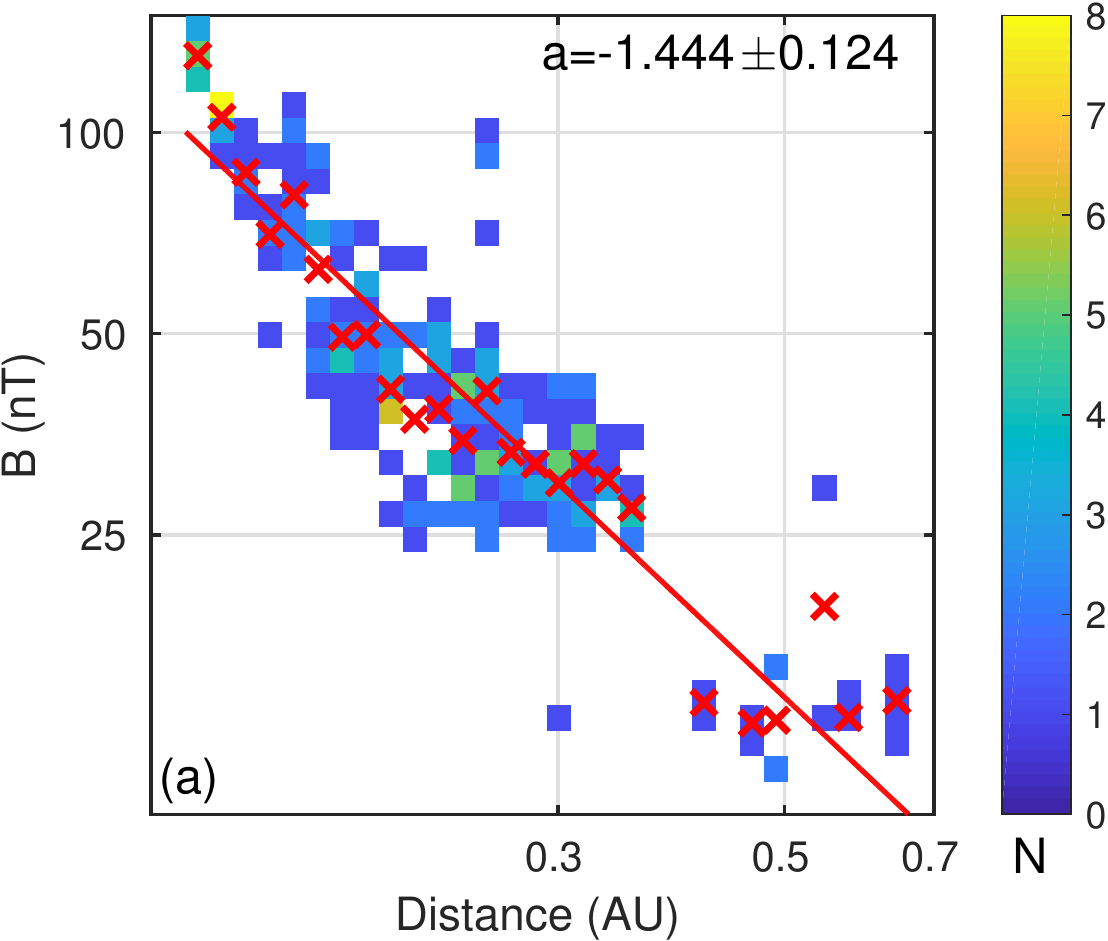}
\includegraphics[width=.3\textwidth]{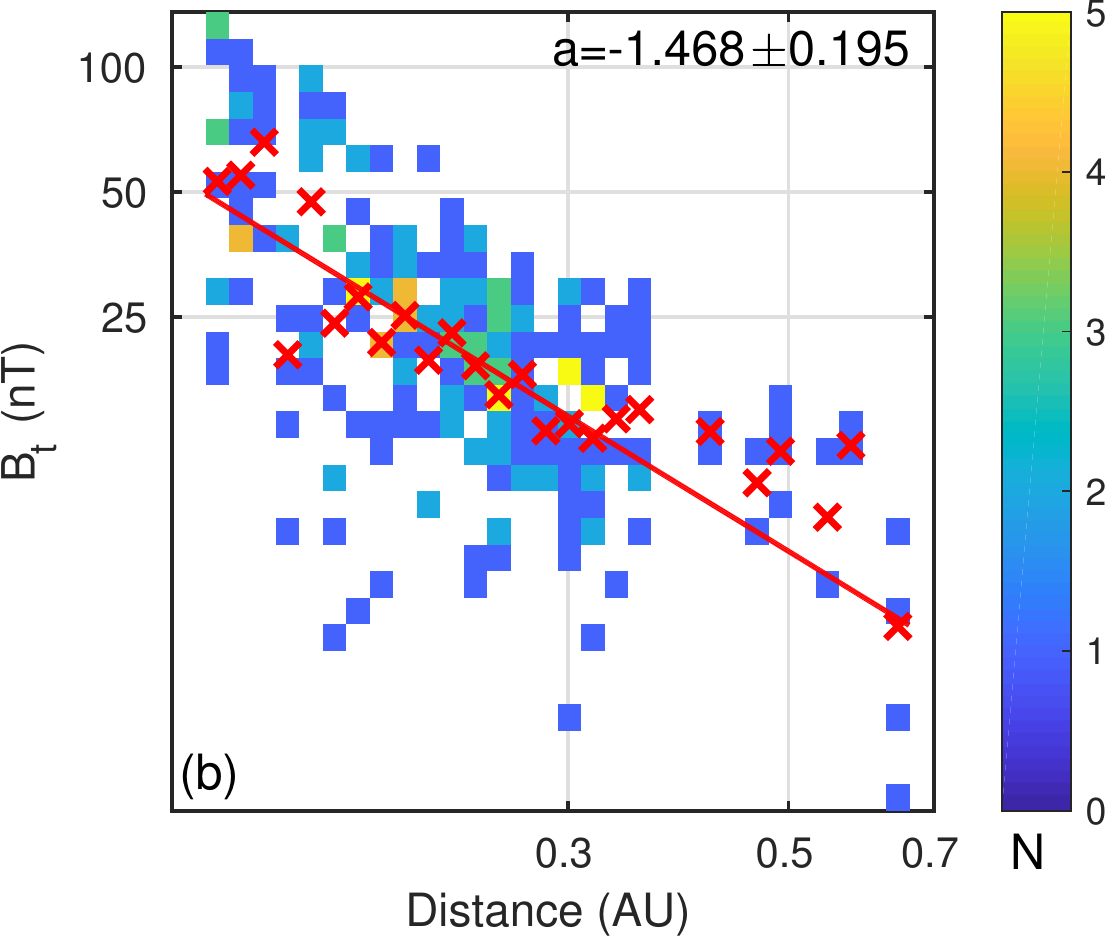}
\includegraphics[width=.3\textwidth]{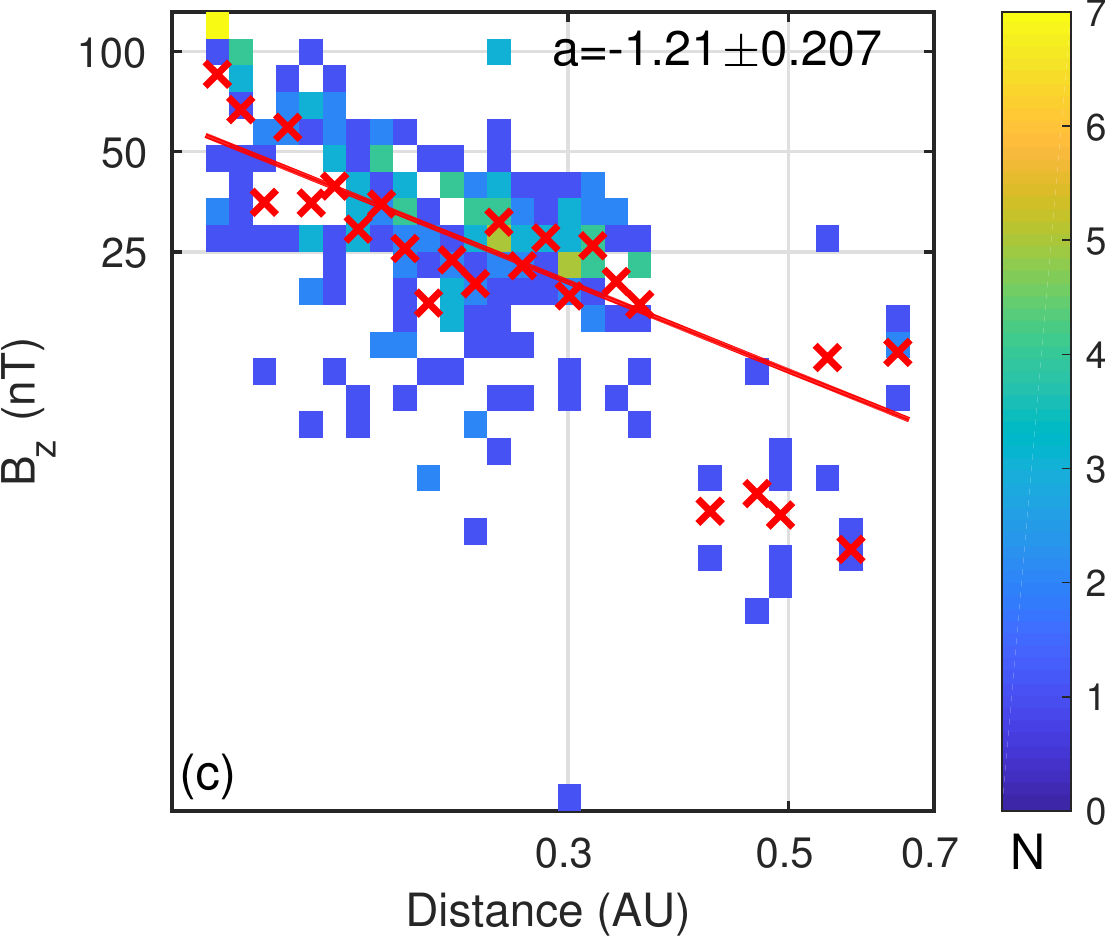}
\includegraphics[width=.3\textwidth]{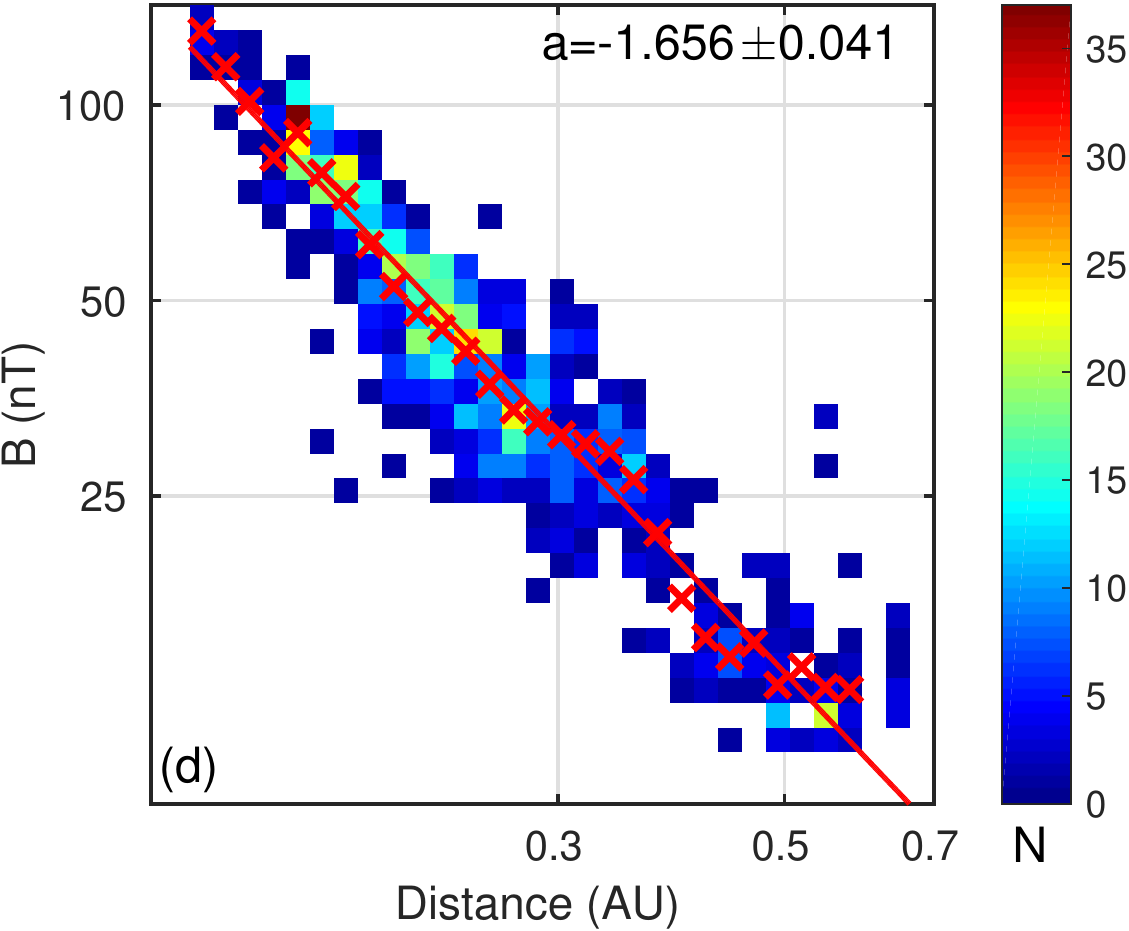}
\includegraphics[width=.3\textwidth]{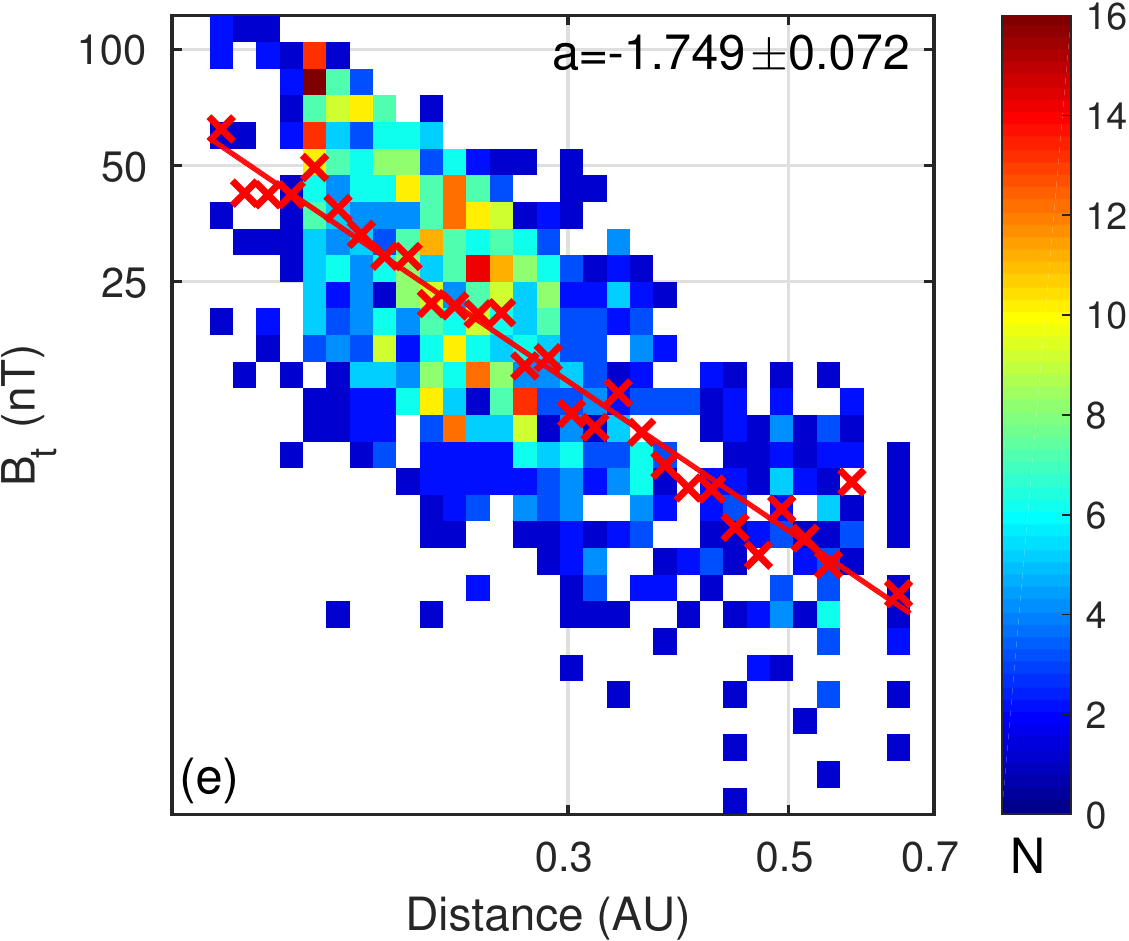}
\includegraphics[width=.3\textwidth]{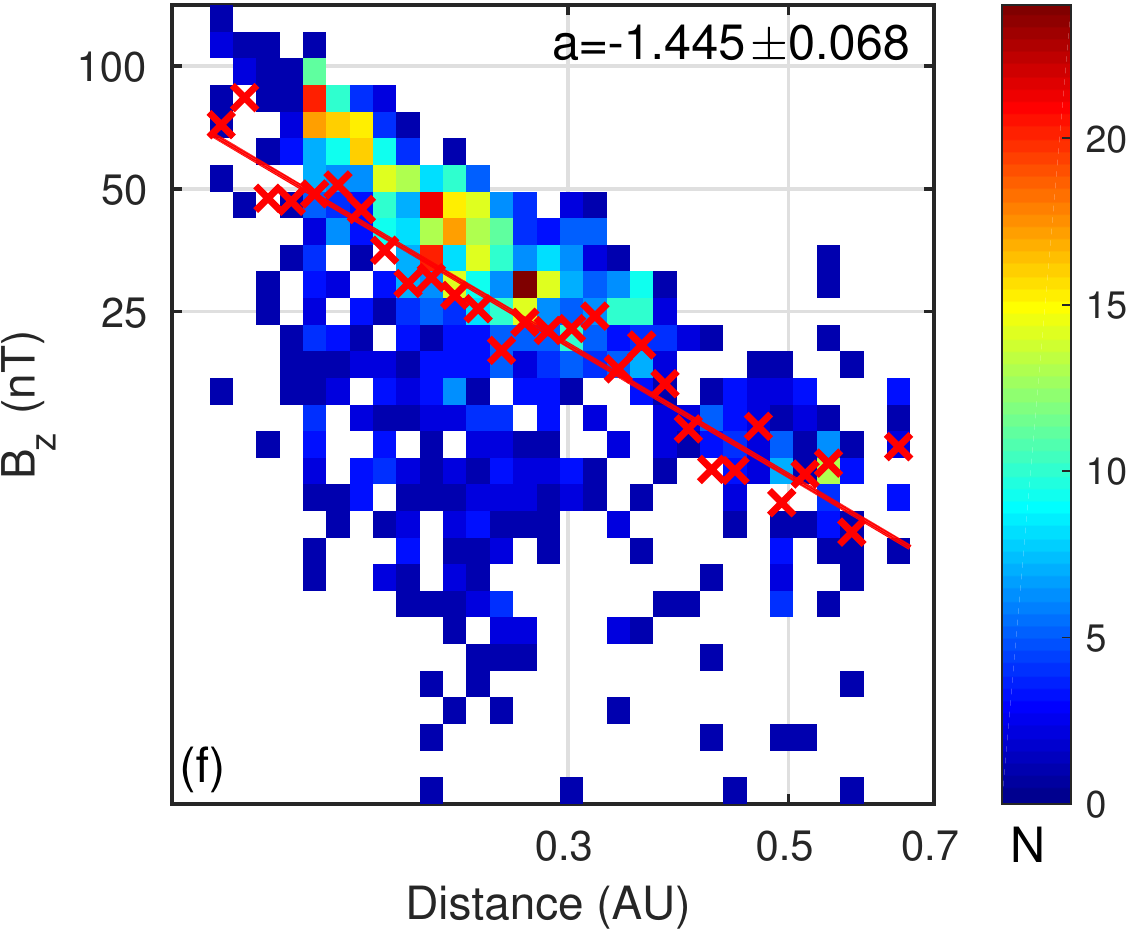}
\caption{The 2D distributions of properties of static FRs (the first row) and FRFFs  (the second row) with respect to the radial distances $r$: (a) \& (d) the average total magnetic field strength, (b) \& (e) the transverse field strength $B_t$, and (c) \& (f) the axial field strength $B_z$. The format follows that of Figure \ref{fig:FRFF1}. The red line on each plot denotes the power-law fitting, with the power-law index and the uncertainty listed in the upper right. }\label{fig:FRFF2}
\end{figure}

In Figure \ref{fig:FRFF2}, we display the radial variation of the averaged magnitude of the total magnetic field $B$, the transverse field $\textbf{B}_t$, and the axial field $B_z$, inside FRs and FRFFs. They generally exhibit tighter decaying trends as guided by the red lines. Figure \ref{fig:FRFF2}(a) \& (d) present the field strength of both FR and FRFF records which tend to fit a power-law change $r^{a}~(a < 0)$, i.e., indicating a radial decay with increasing $r$. The average values in each bin of $r$ follow the power-law functions as indicated by the red lines with indices and associated uncertainties denoted. The indices for groups of FR and FRFF are quite different given the limited sample sizes. Comparing with FRs, the magnetic fields of FRFFs may decay more rapidly with increasing $r$.

\section{Connection with Switchbacks and Selected Case Studies}
\label{sec:switchbacks}

Magnetic field switchbacks refer to the sudden reversals of the radial magnetic field which are accompanied by velocity enhancements. Multiple switchbacks were observed and reported during the first two PSP encounters \citep{Bale2019,Kasper2019,Horbury2020,Dudok2020}. Since both switchbacks and flux ropes are related to magnetic field rotation, one may doubt whether they have any connection. On one hand, the Alfv\'enic fluctuation correlating the magnetic field and flow velocity can possibly exist at FR boundaries \citep{Borovsky2008}. Furthermore, \cite{Pecora2019} found significant correlation between magnetic discontinuities and FR boundaries based on observational analysis at 1 au. Similarly, the macroscopic analysis of switchbacks hints as well that a spacecraft may encounter switchbacks when a spacecraft travels across magnetic flux tubes \citep{Dudok2020}. On the other hand, when taking the apparent Alfv\'encity into account, these switchbacks may be found to coincide with FRFFs, not just near the boundaries. 

\begin{figure}
\centering
\includegraphics[width=.3\textwidth]{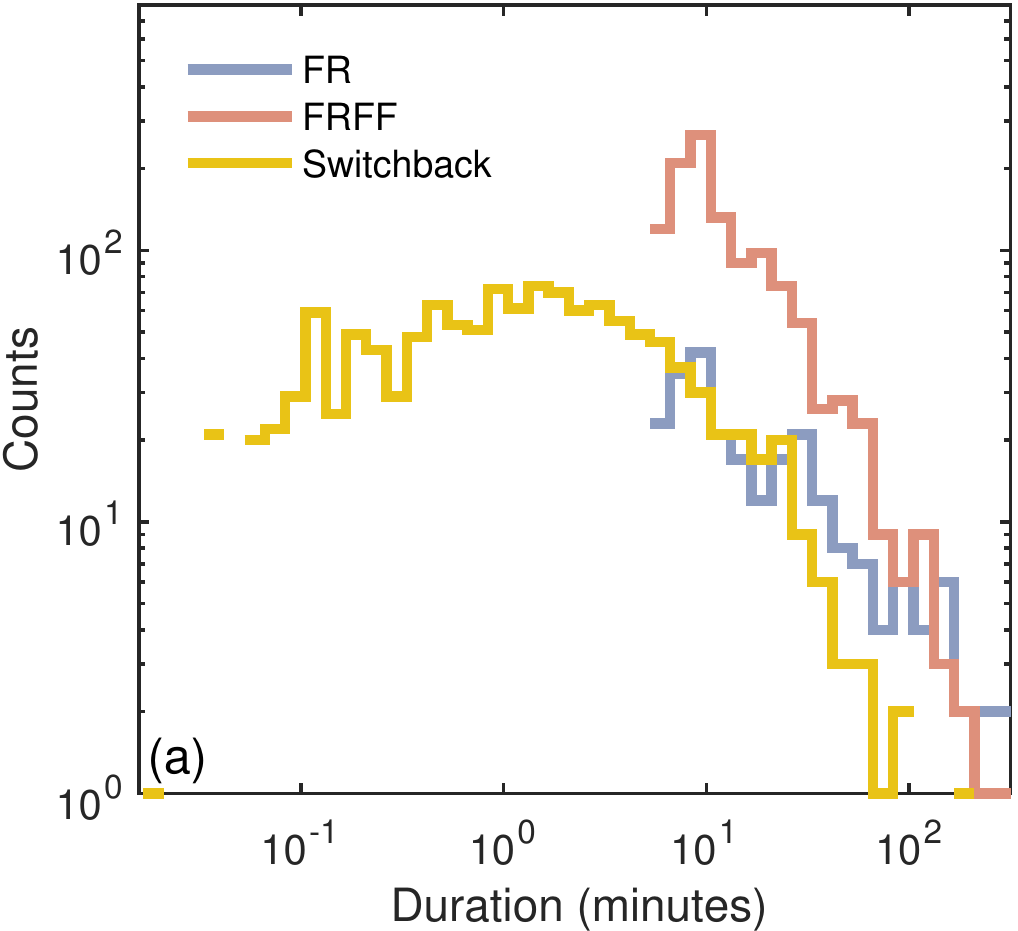}
\includegraphics[width=.3\textwidth]{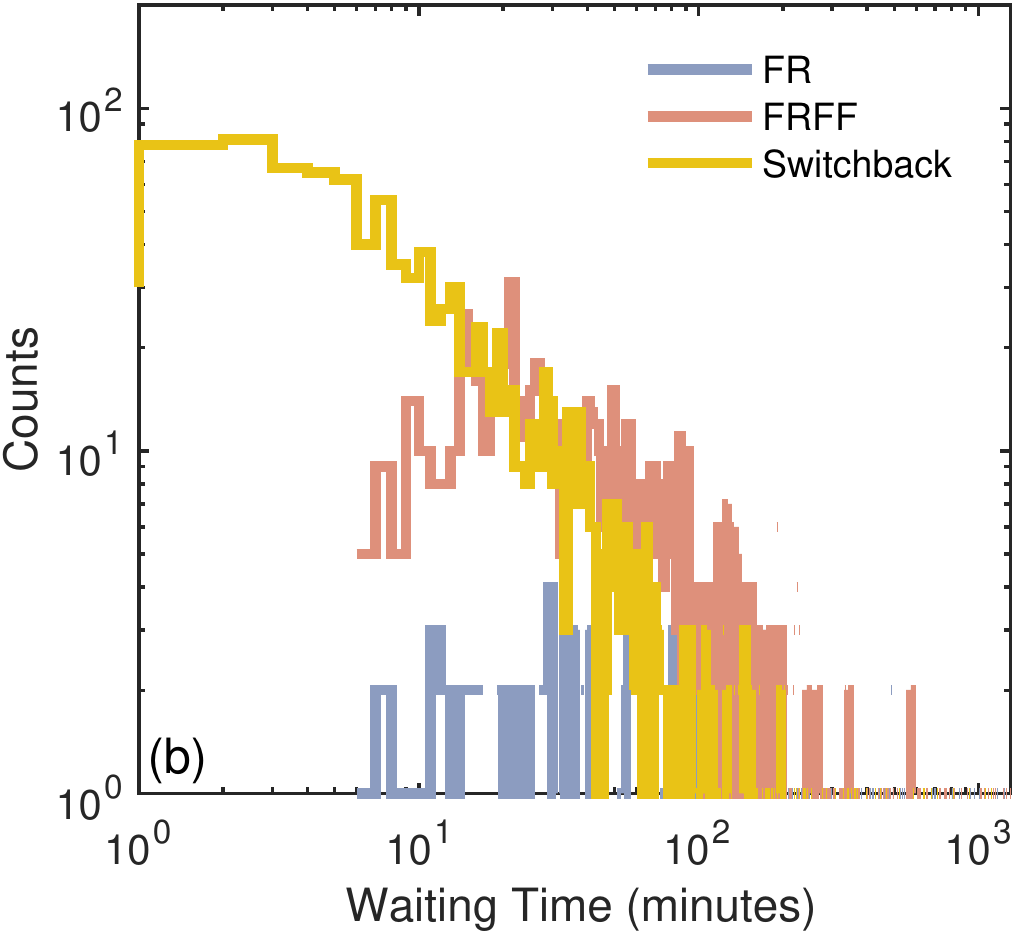}
\includegraphics[width=.3\textwidth]{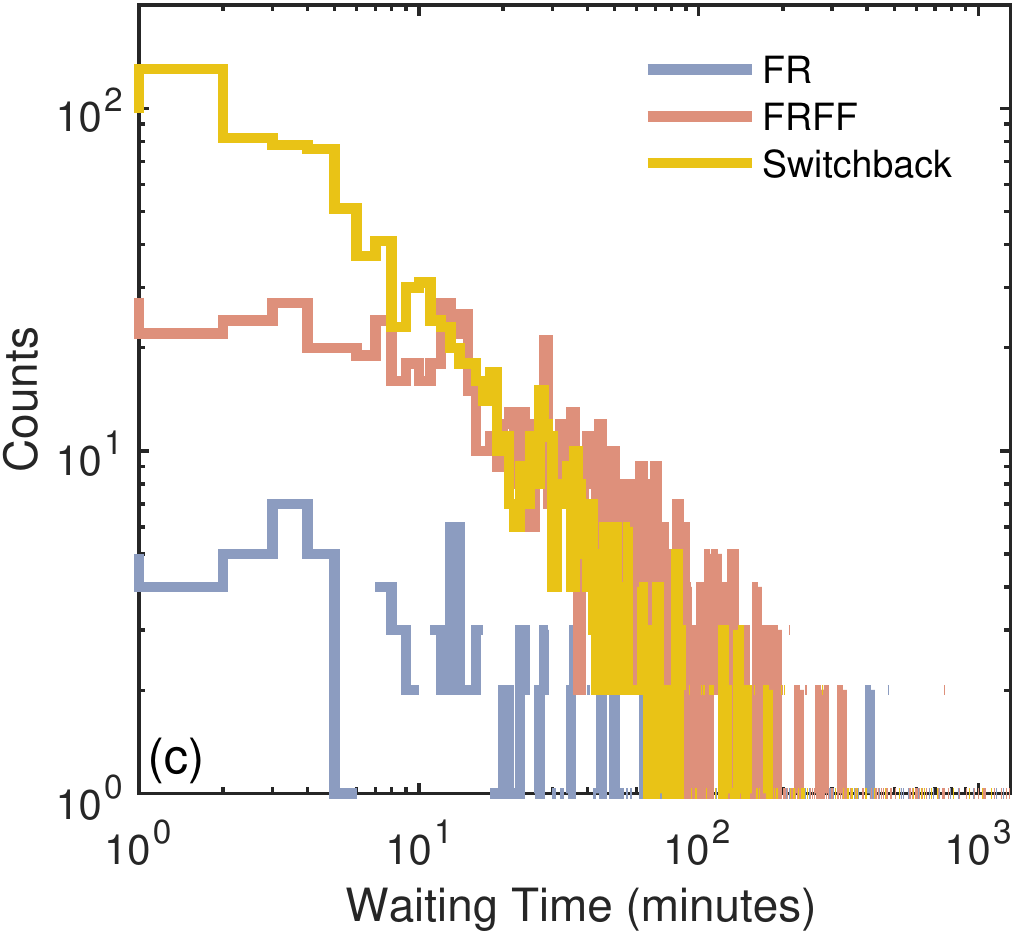}
\caption{Distributions of (a) duration, (b) the first, and (c) second types of waiting time for FRs, FRFFs, and switchbacks. The first and second types of waiting time are respectively defined as $Start_{i+1}-Start_{i}$ and $Start_{i+1}-End_{i}$, where $i$ represents the event sequential number. }\label{fig:wait}
\end{figure}

Given the increased sample size of events from E1 to E5, especially by including FRFFs which share the same property of being Alfv\'enic with switchbacks, we compare the FRFFs and FRs with the velocity spikes in lists of switchbacks \citep{Kasper2019,Martinovic2021}. The identification of switchbacks in their study requires an occurrence of a clear rotation of the magnetic field by more than 45$^\circ$ for over 10 s with respect to the quiet solar wind background and returning to the normal condition. They detected 1235 events in three encounters, E1, E2, and E4. Among them, there are 538 events in E1 from 2018 October 31 to November 10, 383 from 2019 March 30 to April 10, and 314 from 2020 January 24 to February 3. The spike refers to an interval when the simultaneous changes of the magnetic field and velocity happen, without the leading/trailing edges from the quiet period. The duration of spikes (start to its end) ranges from 1 to 9316.96 seconds.

Figure \ref{fig:wait} displays the distributions of duration and two types of waiting time for FRs, FRFFs, and switchback spikes. As shown in Figure \ref{fig:wait}(a), the distributions of duration generally exhibit power laws over similar duration ranges between $\sim$10 and 100 minutes. Spikes were also found to possess rather short durations, while our current detection of FRs and FRFFs has a lower limit on duration at 5.6 minutes. Moreover, the power-law tendencies of FRFFs and spikes are more similar, which seems to hold for the waiting time distributions. Figure \ref{fig:wait}(b) and (c) present the two types of waiting time analyses. The first type measures the time difference between two starting times of adjacent event intervals ($Start_{i+1}-Start_{i}$, where $i$ represents the event sequential number), whereas the second type is calculated by the difference between the starting time of an event and the ending time of the preceding one ($Start_{i+1}-End_{i}$). Notice that the first type of waiting time consists of the event duration, i.e., $End_{i} - Start_{i}$, and the second type of waiting time. The minimum of the first type of waiting time is thus identical to that of duration, i.e., 5.6 min, while the minimum for the second type can extend to zero. 
Again, the waiting time of FRs distributes distinctly from the other two sets, while those of FRFFs and spikes appear similar and exhibit the power-law tendencies, especially at ranges greater than 10 min. Such analogous distributions of waiting time hint that switchbacks and FRFFs may share certain intrinsic processes in their generation and evolution. 

In addition, we calculate the time difference between all events in order to find relative positions between spikes and FRFFs as well as FRs. The term `adjacent event' is tagged when one is recorded next to FRFFs and FRs or neighboring to these events within certain temporal ranges. We find that about 39 out of 1,235 spikes in encounters No.1, 2, and 4 occur near the FRs and FRFFs within $\pm$ 1 min. When relaxing this limit, it becomes 161 in $\pm$ 5 mins and 273 in $\pm$ 10 mins. These numbers suggest that spikes may co-exist with FRFFs and FRs, and have a certain probability that can be detected when a spacecraft passes through a sea of kinked structures.

\begin{table}
\begin{center}
\caption{Case Studies: FR and FRFFs Overlapping with Switchbacks.}
\footnotesize
\begin{tabular}{lccccc}
\toprule
Cases & Time Periods (UT) & Wal\'en Test Slope & $\langle M_A\rangle$ & Spike Periods (UT)	\\
\midrule 
1 & 2018 Nov 2, 12:29:15 - 13:21:03 & 0.28 & 0.3290 & 13:05:18 - 13:14:52\\
2 & 2018 Nov 6, 02:11:27 - 02:18:55 & 0.33 & 0.3533 & 02:07:31 - 02:20:02 \\
3 & 2020 Jan 24, 22:51:12 - 23:01:56 & 0.42 & 0.4278 & 22:57:41 - 23:06:55\\
\bottomrule
\end{tabular}
\label{table:case}
\end{center}
\end{table}

Moreover, there are 74 intervals for which FRFFs or FRs overlap with spikes. An overlapping is tagged when: (1) the starting time of spikes locates within an FRFFs or FRs interval, or (2) the starting time of FRFFs or FRs is enclosed by a spike interval. We select three overlapping cases with identified spikes for reconstructing 2D cross-section maps: one is from the FR group and the other two are from FRFF. Table \ref{table:case} lists the basic information about these three cases including time periods, the absolute values of the Wal\'en test slopes, $\langle M_A\rangle$, and the corresponding spike intervals. Case No.1 starts from 2018 November 2, 12:29:15 to 13:21:03 UT. Values of the Wal\'en test slope and $\langle M_A\rangle$ are quite marginal for a criterion of FRFF event, i.e., 0.28 and 0.33, respectively. The overlapping spike is identified on the same day from 13:05:18 - 13:14:52 UT. Cases No.2 and No.3 are records from the FRFF group, which start from 2018 November 6, 02:11:27 to 02:18:55 UT and 2020 January 24, 22:51:12 to 23:01:56 UT, respectively. These two cases possess relatively modest Alfv\'enicity. The corresponding spikes are reported in \cite{Kasper2019} and included in a recently extended list \citep{Martinovic2021}.
 
\begin{figure}
\centering
\includegraphics[width=1.0\textwidth]{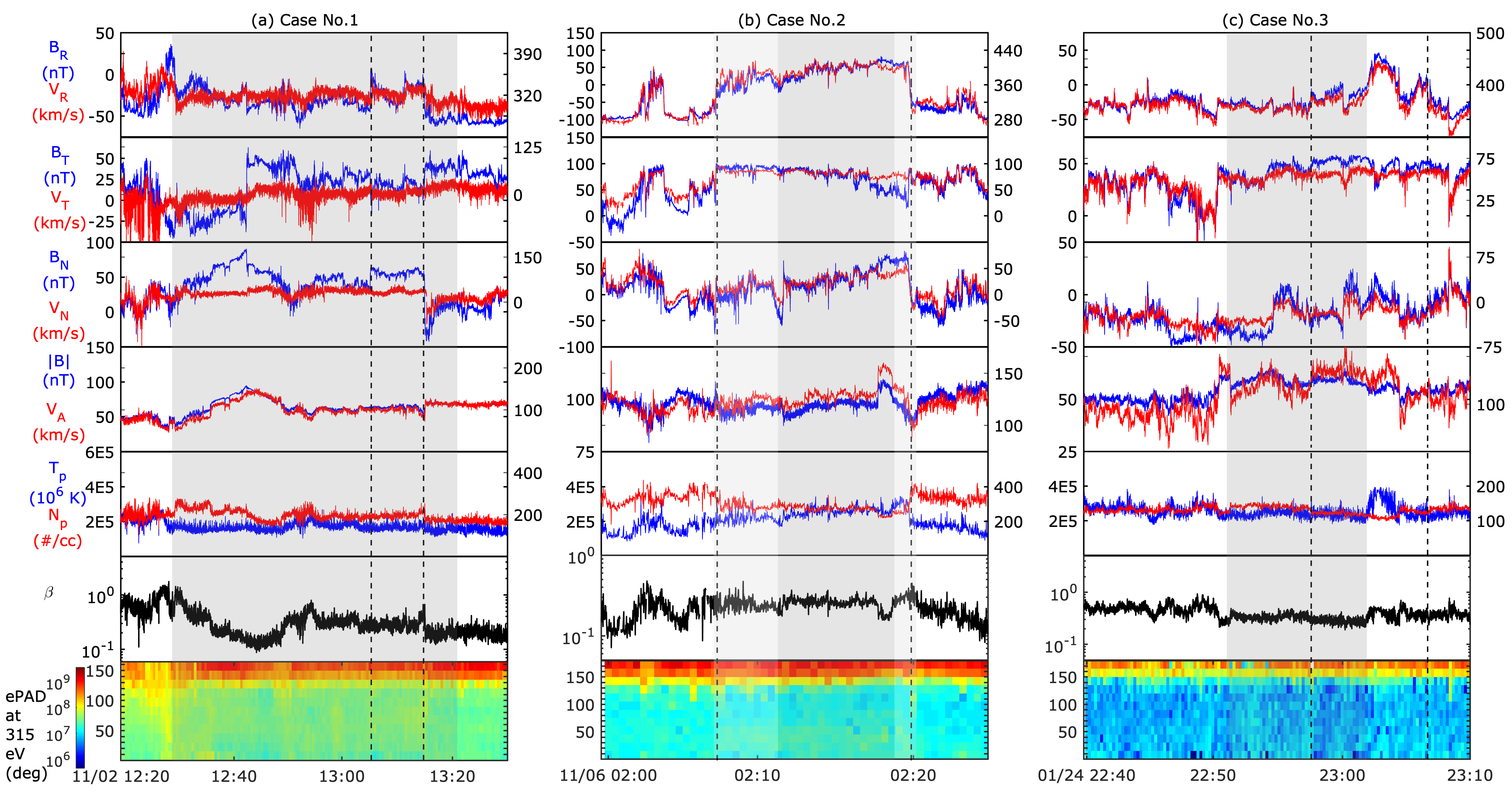}
\caption{Time-series plots for cases listed in Table \ref{table:case}: (a) No.1, (b) No.2, and (c) No.3. Panels from the top to the bottom are: three components of the magnetic field and solar wind bulk velocity $B_R$ \& $V_R$, $B_T$ \& $V_T$, $B_N$ \& $V_N$, the field magnitude $|B|$ and the Alfv\'en speed $V_A$, $T_p$ and $N_p$,  plasma $\beta$, and ePAD at 315 eV. The left y-axis describes parameters in blue, while the right y-axis describes parameters in red. The FRFF/FR events are marked by gray areas, while the boundaries of a spike are denoted by two vertical dashed lines. The light gray areas in panel (b) mark the extended portion for the purpose of reconstruction. }\label{fig:casetms}
\end{figure}

In Figure \ref{fig:casetms}, we present the time-series variations around these three cases. The FR/FRFF events and spikes are denoted by gray shading areas and vertical dashed lines, respectively. The relative locations of FR/FRFF events and spikes have three situations corresponding to the sub-figures, respectively: (a) the spike is fully enclosed by an FR; (b) both boundaries of a spike enclose those of an FRFF event; (c) the FRFF overlaps with a part of a spike. Reversals of the radial magnetic field $B_R$ between two vertical lines are obvious for all three events. The variations of $N_p$ and $T_p$ at boundaries of spikes are different. For example, they all have a sudden drop at the trailing edge in Figure \ref{fig:casetms}(a). Although density $N_p$ drops in Figure \ref{fig:casetms}(b), the temperature $T_p$ has an inverse change. The variations of these two parameters at those boundaries in Figure \ref{fig:casetms}(c) are very limited. On the other hand, within an FR interval, both $B_R$ and $B_T$ have evident bipolar rotations from the negative values to the positive or vice versa, whereas it happens mainly in the $N$ direction within intervals of the two FRFFs. As aforementioned, case No.1 is an FR event. Comparing with the two FRFF events which have three velocity components align and corotate with the magnetic field for almost the whole interval, bulk velocity inside an FR may largely lack or only have this correspondence for a short time period. Therefore, this transient Alfv\'enicity, although existed, will be suppressed by the low-Alfv\'enicity of the whole interval. The last panel is the corresponding ePAD distribution for each time period. Directions of electron strahls remain unchanged before and after the PSP traversings of the FR, FRFF, and switchbacks.

\begin{figure}
\centering
\includegraphics[width=0.8\textwidth]{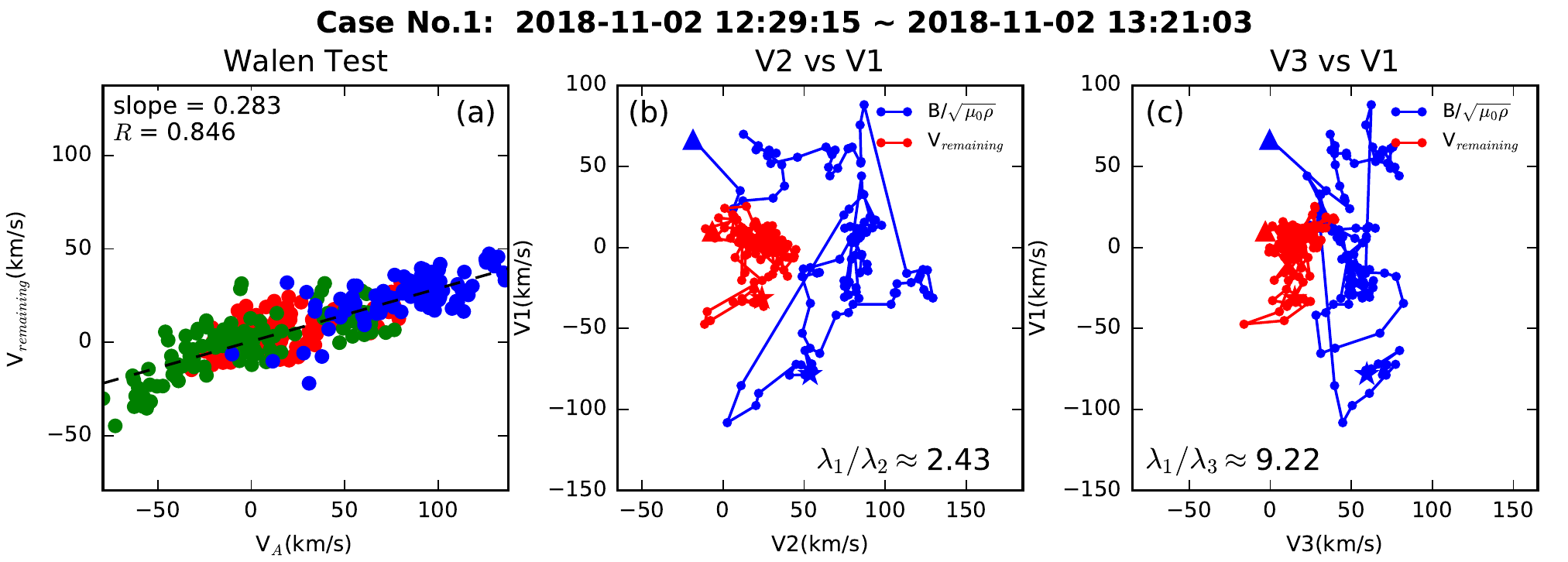}
\includegraphics[width=0.8\textwidth]{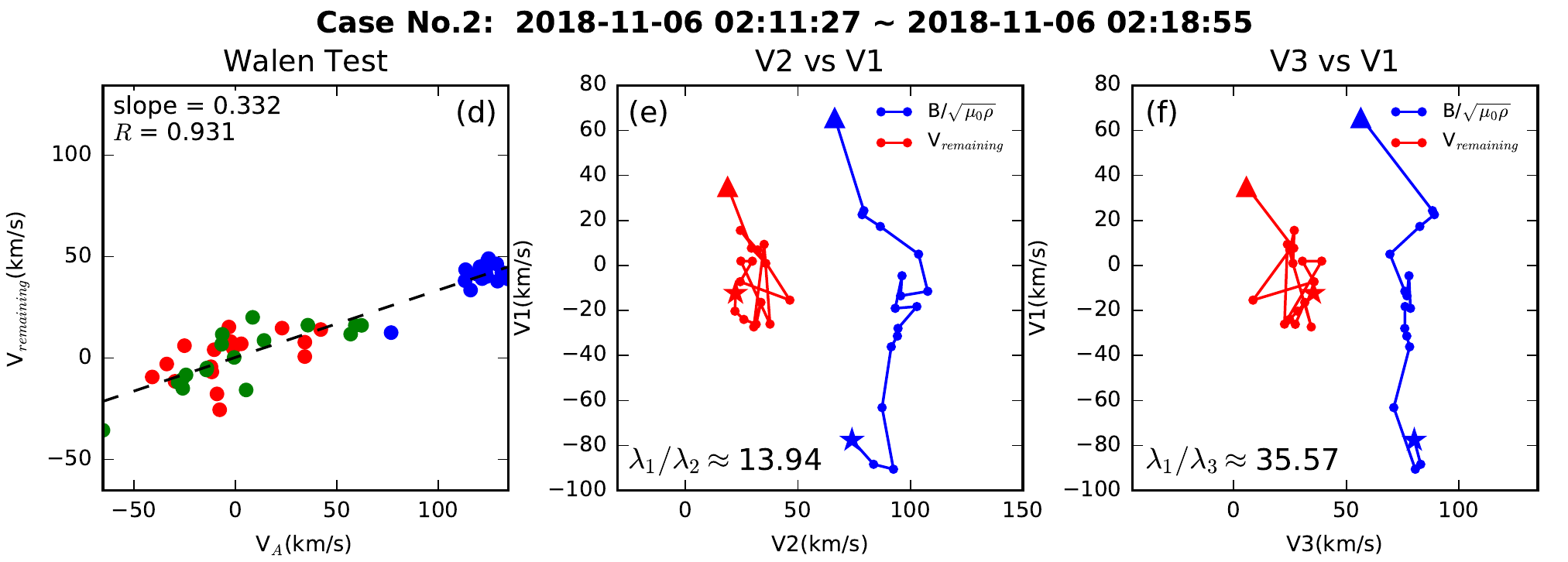}
\includegraphics[width=0.8\textwidth]{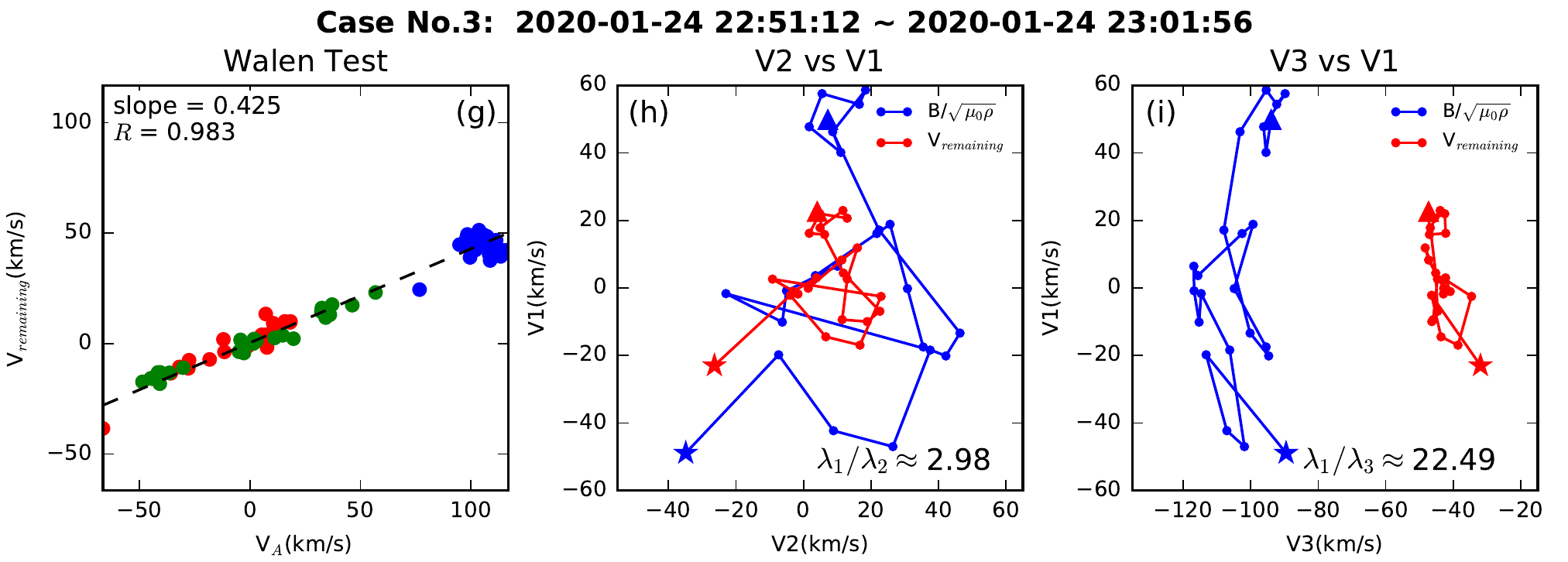}
\caption{The Wal\'en relation and hodograms for the cases listed in Table \ref{table:case}: (a-c) No.1, (d-f) No.2, and (g-i) No.3. Panels from the left to right are the Wal\'en relation in the co-moving frame with three components $(\hat{\mathbf{x}},\hat{\mathbf{y}},\hat{\mathbf{z}})$ and hodograms of the velocities. The dashed line in the 1st column represents the linear regression line. $R$ represents the correlation coefficient between two velocities. The parameters $\lambda_1$, $\lambda_2$, and $\lambda_3$ represent the eigenvalues corresponding to the maximum, intermediate, and minimum variance in the magnetic field. The starting and ending points are denoted by triangle and star respectively. }\label{fig:caserl}
\end{figure}

In Figure \ref{fig:caserl}, the 1st column exhibits the Wal\'en relation plot, which shows the linear regression line between the remaining flow velocity $\textbf{V}_{remaining}$ in the frame of reference and the local $\textbf{V}_A$. The three components of velocities are marked in red, green, and blue, respectively. The second and third columns in Figure \ref{fig:caserl} present the hodograms of $\textbf{V}_{remaining}$ and $\textbf{B}/\sqrt{\mu_0 \rho}$, in which the movements of the endpoints of each vector are displayed in a coordinate determined by the principal axes from the corresponding minimum variance analysis of the magnetic field (MVAB, \cite{Sonnerup1998}). The subscripts 1, 2, and 3 represent the directions and components of the maximum, intermediate, and minimum variance. The magnetic field for all three cases shows certain degree of rotation on the hodograms. However, the existence of the flux rope-like structures is better confirmed by the GS reconstruction results. In addition, the variations of endpoints of the remaining flow are different. In case No.1, the similarity between $\textbf{V}_{remaining}$ and $\textbf{V}_A$ is not pronounced, which can also be demonstrated by a small Wal\'en test slope. Cases No.2 and 3 show progressively improved in-sync variation between velocity and magnetic field, an indication that the Alfv\'enicity becomes more pronounced.

\begin{figure}
\centering
\includegraphics[width=.3\textwidth]{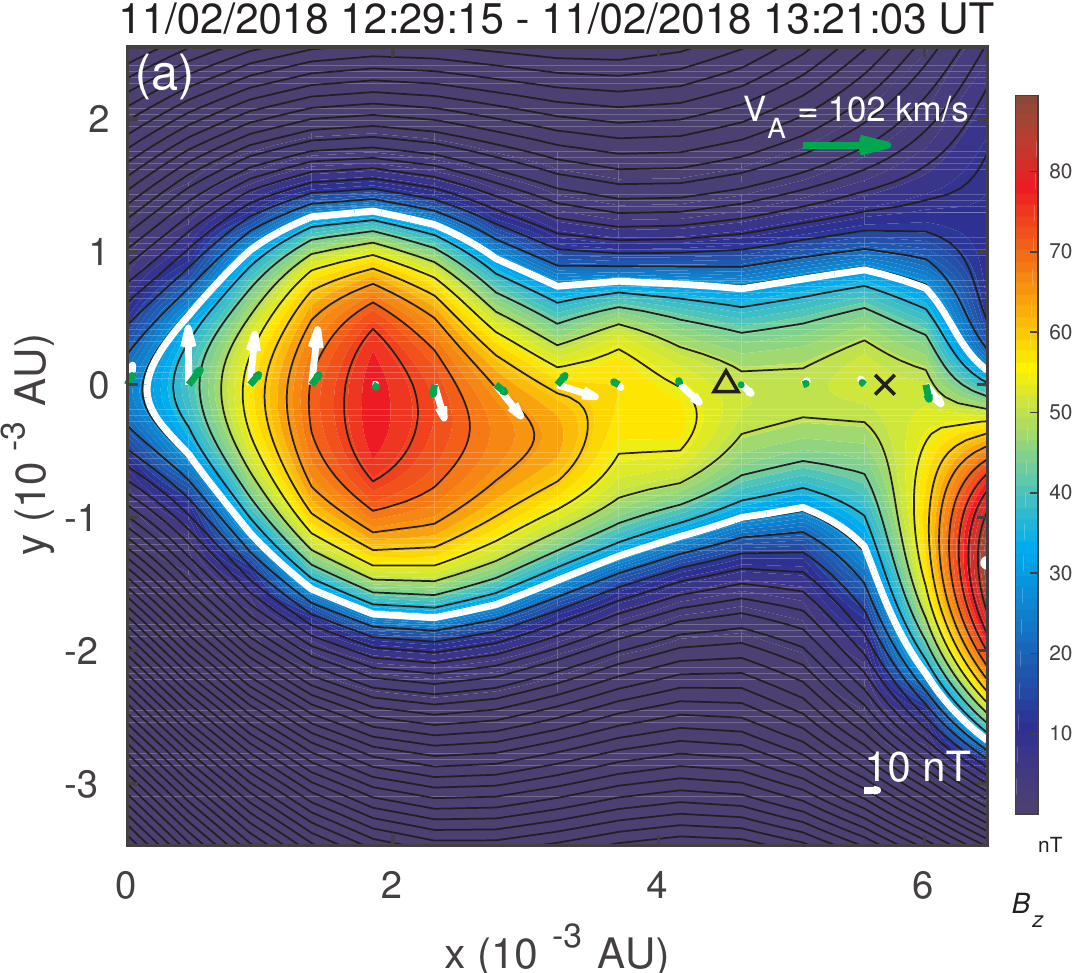}
\includegraphics[width=.3\textwidth]{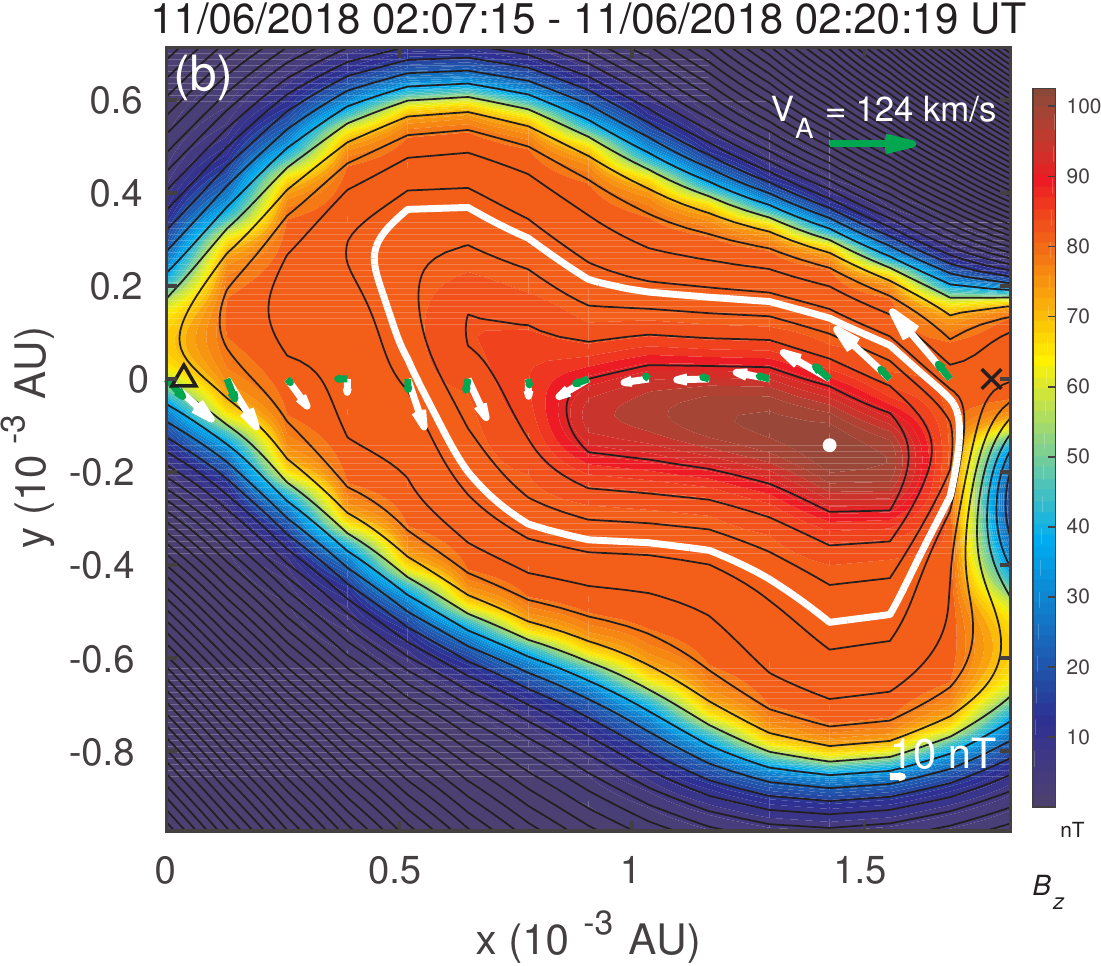}
\includegraphics[width=.3\textwidth]{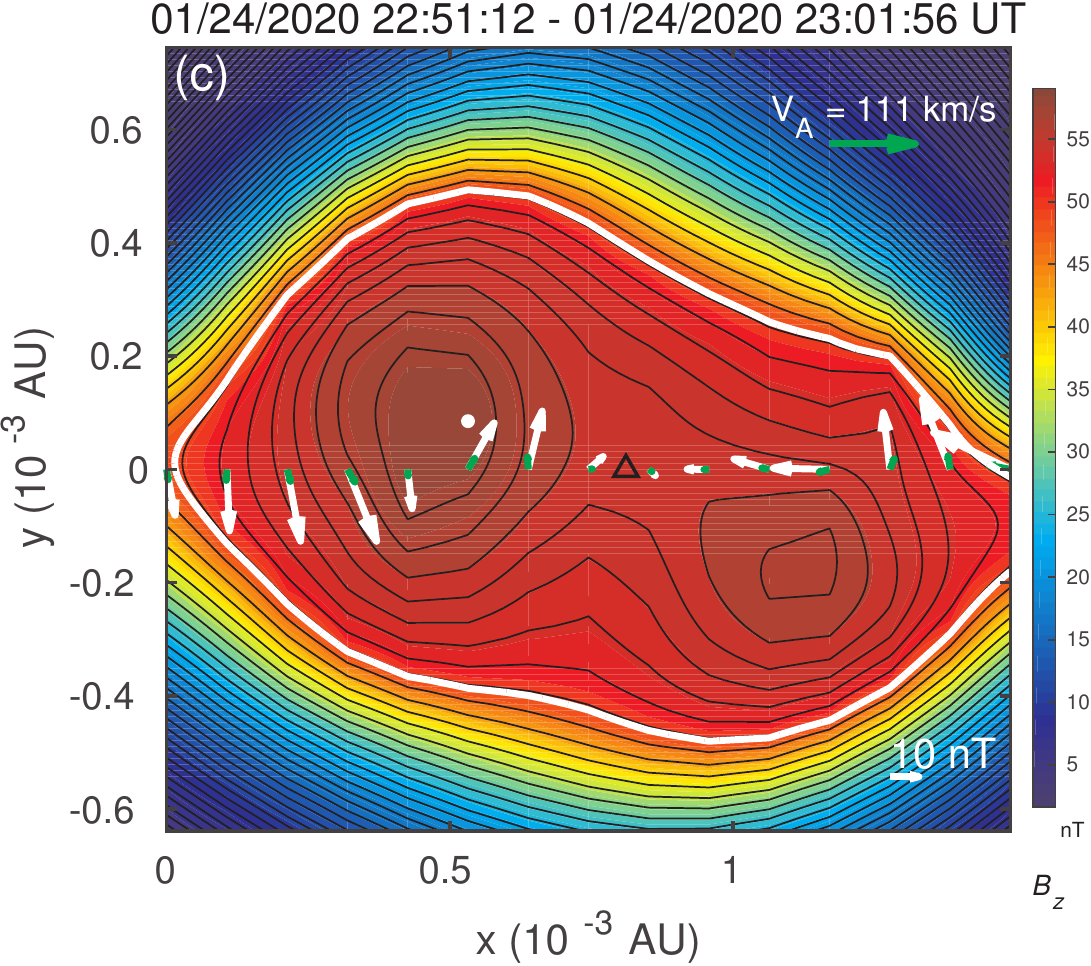}
\includegraphics[width=.3\textwidth]{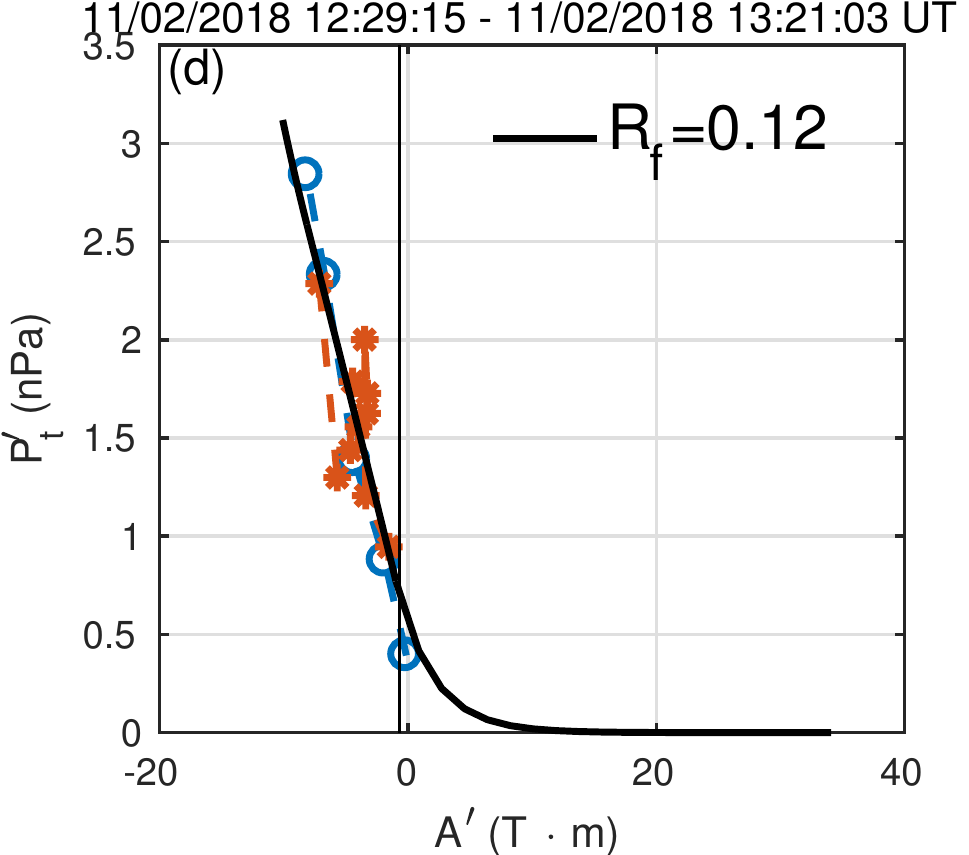}
\includegraphics[width=.3\textwidth]{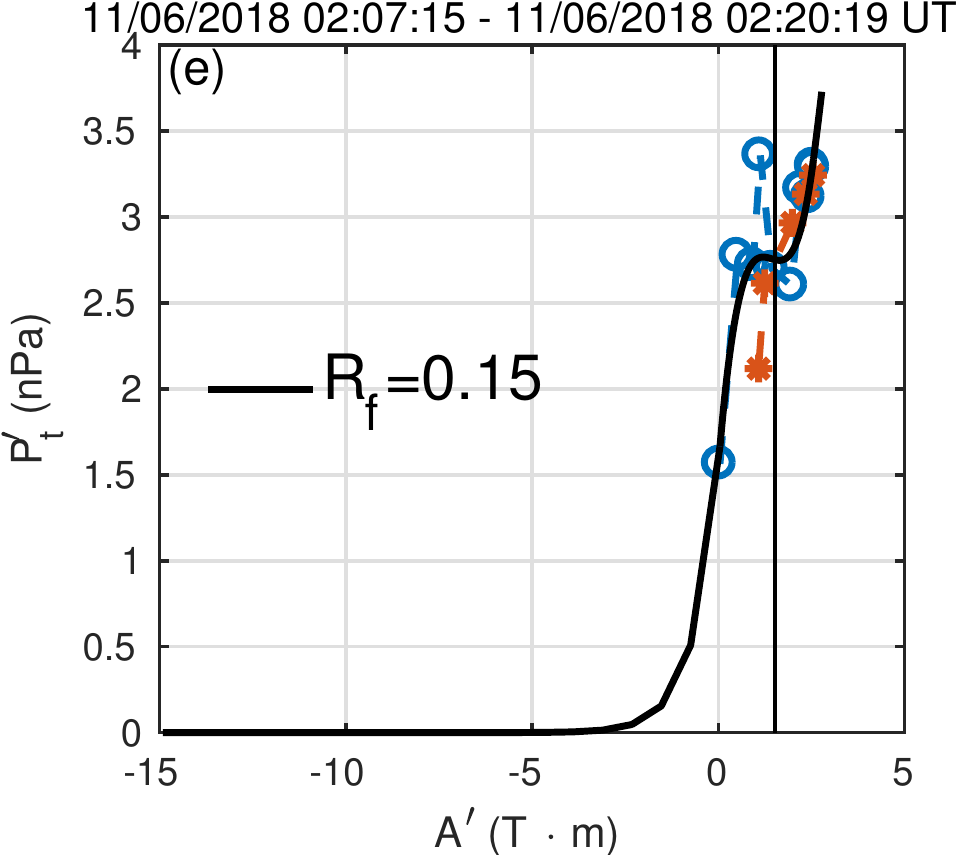}
\includegraphics[width=.3\textwidth]{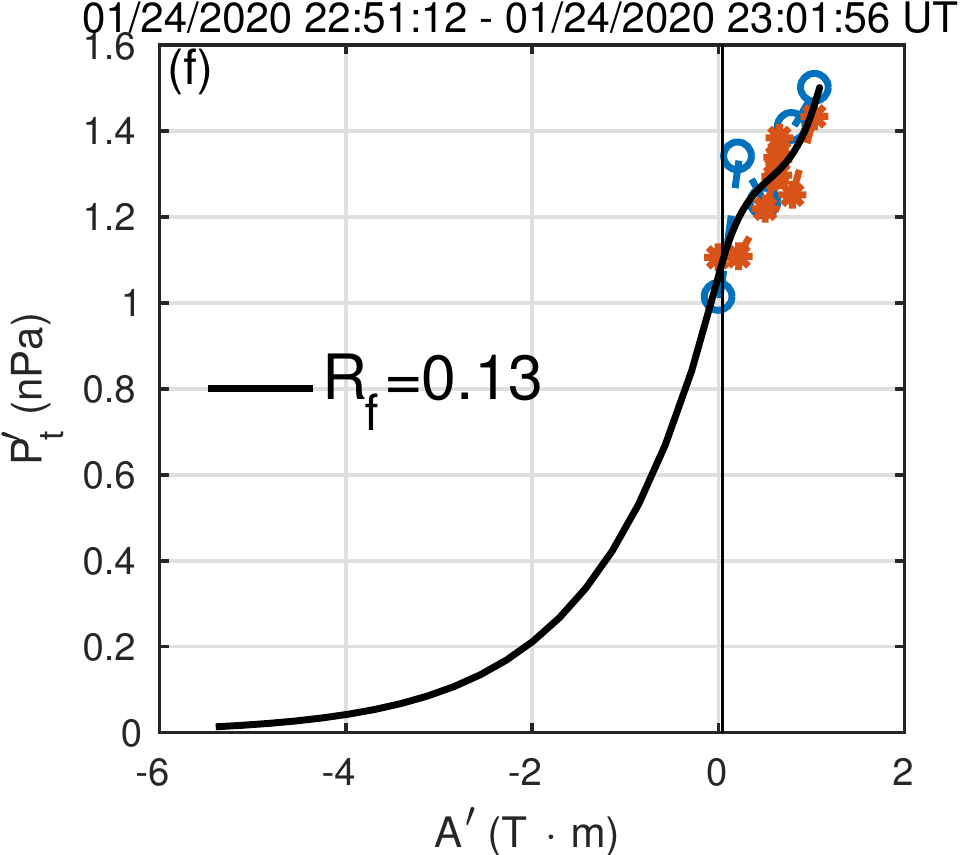}
\caption{Cross-section maps and the $P'_t$ versus $A'$ curves for the cases listed in Table \ref{table:case}: (a) \& (d), No.1, with $\hat{\mathbf{z}}=[-0.383 ,0.321, 0.866]$ in the RTN coordinates; (b) \& (e), No.2 for a modified time interval, with $\hat{\mathbf{z}}=[0.269, 0.899, 0.343]$ in the RTN coordinates;  (c) \& (f): No.3, with $\hat{\mathbf{z}}=[-0.470, 0.813, -0.342]$ in the RTN coordinates. The first row: the black and color-filled contours represent the transverse magnetic field lines $\textbf{B}_t$, and the axial magnetic field $B_z$ as indicated by the colorbar. The white dot, white, and green arrows along $y=0$ denote the maximum  $B_z$, the  $B_t$ and $V_t$ vectors along the spacecraft path, respectively. The triangle and cross symbols mark the boundaries of the overlapping spike record. The second row: the $P'_t$ versus $A'$ measurements (symbols) and the corresponding functional fits (black curves) for the two branches with the double-folding pattern with the fitting residue $R_f$ denoted. The vertical line denotes a boundary of certain magnetic flux function value, corresponding to the white contour line in the top panel.}\label{fig:map}
\end{figure}

By using the new GS type reconstruction described in Section \ref{sec:method} \citep{Teh2018}, we recover the 2D cross-sections of these three cases, shown in Figure \ref{fig:map}. In each top panel, the black contours and color background describe the transverse $\textbf{B}_t$ and the axial field $B_z$, which form the closed transverse field-line regions, i.e., indicative of magnetic flux rope configuration. Vectors of the remaining flow and the transverse magnetic field $\textbf{B}_t$ along the spacecraft path are denoted by green and white arrows along $y=0$. The starting and ending times of switchbacks are marked by triangles and crosses, respectively, along $y=0$. The bottom panels of Figure \ref{fig:map} are the corresponding $P'_t$ versus $A'$ curves from which the cross-section maps are reconstructed.

As aforementioned, the Alfv\'enicity of case No.1 is not strong and thus the flow speed is much smaller than the local Alfv\'en speed. The shape of this event is irregular, which may be ascribed to a sharp transition of $B_T$ from $\sim$ -30 to 50 nT in Figure \ref{fig:casetms}(a). The overall picture possibly consists of a section with a magnetic flux rope in an elongated current sheet. The boundaries of the corresponding spike, 13:05:18 - 13:14:52 UT, are denoted by a triangle and a cross on Figure \ref{fig:map}(a). This spike interval appears to locate inside a current sheet, possibly enclosing an X-point configuration prone to the magnetic reconnection. In fact, \cite{Froment2021} reported the existence of the reconnection exhausts at both boundaries of a spike from 13:05:09 to 13:15:11 UT that is nearly identical to the one we use here. 

Figure \ref{fig:map}(b) is the modified case No.2. As listed in Table \ref{table:case}, this FRFF event is enclosed by the spike, and both boundaries of the FRFF are near those of the spike. Therefore, we manually extend the reconstruction interval in order to present the full structure. As shown by light gray areas in Figure \ref{fig:casetms}(b), the left and right boundaries are extended from the original FRFF to spike, i.e., 02:11:27 to 02:07:15 UT and 02:18:55 to 02:20:19 UT, respectively. Directions of the magnetic field at starting and ending times are completely reversed. Also, the plasma flows are aligned with the local magnetic field, especially at boundaries. The center of these closed field lines is the core region of the FRFF event, typical of a flux rope configuration. Also, one should notice that the direction of the magnetic field from the start of spike (triangle symbol) to the main structure of FRFF (around the fifth white arrow from the left) is almost unchanged.

Figure \ref{fig:map}(c)\&(f) depict an FRFF map that overlaps with a part of the spike. The two branches of $P'_t$ versus $A'$ curves fold onto each other very well and thus enables the recovering of the single and complete 2D cross-section. The leading boundary of the spike coincides with the X-point like structure in the middle, characteristic of or a remnant of possible magnetic reconnection site. 

\section{Summary and Discussion}\label{sec:sum}

In summary, we report the FR and FRFF events via the GS-based algorithms during the first five PSP encounters and present the preliminary analysis of the connections between FRs/FRFFs and switchbacks. In this study, 243 FRs and 1,153 FRFFs are recorded by the PSP at heliocentric distances from 0.13 to 0.66 au in about 4 months. Specifically, we present the overview plot around the fifth perihelion from 2020 May 27 to June 18. Those FR and FRFF events are categorized into two individual groups and analyzed via simple statistics and 2D histograms. Namely, the distributions of duration, scale size, average solar wind speed, and FR/FRFF magnetic fields versus radial distances $r$, are shown. Moreover, we compare the FR/FRFF events with the switchback spikes in \cite{Kasper2019,Martinovic2021}. Three cases are selected and displayed with their time-series plots, cross-section maps, and the $P'_t~versus~A'$ curves, derived from the GS-type reconstruction. In addition, the Wal\'en relations, and hodograms for the remaining flow velocity and the Alfv\'en velocity $\textbf{V}_A$, are shown to illustrate the respective levels of Alfv\'enicity for the three cases. 

The main findings are summarized below.

\begin{enumerate}

\item The event count of FRFF is generally more than that of FR during the first five encounters. Since the Alfv\'enic structures prevail during the E1-E3, the ratio of the counts between FRFFs and FRs jumps up to 17. More FRs occurred during the E4 due to the two HCS crossings, while the number of FRFFs reduces a little due to data gaps around the perihelion.

\item The E5 has distinct results from the others, during which the event counts for the two sets are quite comparable, i.e., 119 for FRs and 148 for FRFFs, respectively. Based on the variation of the radial magnetic field $B_R$, it may be concluded that the PSP may have crossed the HCS around 2020 June 8, which results in the occurrence of FRs, especially preceding the main crossing period.

\item FR and FRFF events are separated by the Wal\'en test slope threshold value 0.3, which is nearly equivalent to the average Alfv\'enic Mach number $\langle M_A\rangle$ of the same value. Most FRFF events possess modest sub-Alfv\'enic flow aligned with the magnetic field. The distributions of the plasma $\beta$ for both FR and FRFF events peak at the same value. Also, both the polar angle $\theta$ and the azimuthal angle $\phi$ of the $z$-axis orientations have analogous distributions for both event sets. 

\item The average duration and scale size, from our current samples, have little variation with the heliocentric distances $r\in[0.1,0.7]$ au from the Sun. Static FRs occur more often in the slow solar wind closer to the Sun, while 33\% of FRFFs  exist in the medium and fast-speed wind. The magnetic fields of both sets of events have clear decaying relations with the increasing radial distances $r$.

\item The distributions of duration and the waiting time analyses of FRFFs, FRs, and switchback spikes present general power-law decaying tendencies. Although these tendencies for the group of FRs are less pronounced due to the small sample size, those of FRFFs and spikes are quite analogous, especially at ranges from $\sim$10 to 100 minutes. 

\item Seventy-four FRFF/FR intervals are found to overlap with those of spikes.  The cross-section maps of the selected case studies indicate the co-location of the flux rope structures with spikes. The topological features revealed by the cross-section maps via the new GS type reconstruction, such as the elongated current sheet, and the embedded X-point, coincide with the identified spikes. Some are accompanied by magnetic reconnection signatures. 

\end{enumerate}
The distributions of properties are similar for FRs and FRFFs based on the limited sample sizes. The small $\beta$ values dominate for both sets, which indicates that most events are likely characterized by the dominance of magnetic field over thermal pressure. The magnetic fields exhibit consistent power-law decaying relations with respect to $r$ but with slightly different power-law indices. In other words, the FRFF is still akin to the FR in which the magnetic field plays an important role. Notice that the detection for FRFFs is simplified by considering the magnetic pressure only in the present study. But the new GS-type reconstruction is not affected by the detection results since it works individually. The implementation of the full algorithm will be an extension to be included in future studies.

Intrinsically, FRFFs and FRs may have direct connections. \cite{Drake2020} suggested that the flows may exist along the magnetic field within a flux rope. Such a state may result due to the fact that the perpendicular flows decay while the transverse flows remain and align with the magnetic field. On the other hand, there may exist a process for the FRFF to evolve into a static FR due to the radially diminishing Alfv\'enicity, especially in the slow wind. Although there exists evidence that structure with high Alfv\'enicity can occur at distant places in high latitudes \citep{McComas2000}, a well-known fact is that both the cross helicity and the Alfv\'en ratio decrease as the structure moves farther out \citep{Marsch1990,Roberts1987}. Moreover, the striking difference of FR event counts between smaller heliocentric distances and 1 au suggests that there may have supplementary sources in addition to those directly formed at solar corona \citep{Chen2020b}. The evolution from the FRFF to the static FR is possibly one of them. The reduction of the Alfv\'enicity is suggested to be associated with velocity shear that can produce non-Alfv\'enic smaller-scale fluctuations \citep{Roberts1992,Parashar2020}.
 
The velocity spikes and the magnetic field reversals sometimes correspond to small structures, which last for a few seconds to several minutes. Lots of switchbacks have duration less than 1 minute. In this study, many switchback spikes are enclosed within FRFF/FR intervals or locate at one boundary. Case No.2 is a rare event that the spike nearly overlaps with FRFF at both boundaries. In brief, our observational result proves the co-existence of FRFF/FR and switchback. Overlapping and adjacent events further suggest that switchbacks can be detected when a spacecraft passes over a sea of kinked flux rope-like structures. Moreover, similar waiting time distributions and statistics hint that these events may share the common process of generation and evolution. Actually, several mechanisms of how and what switchbacks are generated have been put forward, e.g., the interchange reconnection between open and closed field lines \citep{Bale2019}. This suggestion is further proposed in \cite{Drake2020} that the flux rope can be formed by such interchange reconnection and survive in the unidirectional magnetic field with reconnection being suppressed when these structures propagate outwards. Under this scenario, the magnetic field reversals typical of switchback characteristics can also be recognized when the whole flux rope structure is traversed by a spacecraft. Although their simulation results apply to flux ropes of much smaller scales, the implications of their results may be applicable to the FR/FRFF events presented in this study as well. A further comparison with simulation will be carried out in future work, which may enable us to have a better understanding of the correlation among FRs, FRFFs, and switchbacks. Since the structures in our study are fairly small and the spacecraft are usually separated by relatively large distances, and each can only provide one-point in-situ measurements, it is hard to trace the evolution of a single structure. Therefore, from the observational perspective, we may still continue to pursue statistical analysis as we attempt here as initial steps in order to investigate the inter-connection among these small-scale structures.  

\acknowledgments Y.C. and Q.H. acknowledge NASA grants 80NSSC21K0003, 80NSSC19K0276, 80NSSC18K0622, and NSF grant AGS-1650854 for support. The PSP data are provided by the NASA CDAWeb ({\url{https://cdaweb.gsfc.nasa.gov/index.html/}}). This work is carried out with the help of the Bladerunner cluster in the University of Alabama in Huntsville. In addition, this work was made possible in part by a grant of high performance computing resources and technical support from the Alabama Supercomputer Authority. 

\bibliography{bib_database}

\end{document}